\documentclass[review,12pt]{elsarticle}

\usepackage{amssymb}
\usepackage{amsthm}
\usepackage{framed} 
\usepackage{amsmath,color}
\usepackage{mathrsfs}
\usepackage{graphicx}
\usepackage{epstopdf}
\usepackage{float}
\usepackage{caption}
\usepackage{subcaption}
\usepackage{bm}
\usepackage{bbm}
\usepackage{mathrsfs}
\usepackage{cleveref}
\usepackage{soul}
\usepackage{accents}
\usepackage{color,soul} 
\usepackage{color} 
\usepackage{nomencl} 
\makenomenclature 
\biboptions{sort&compress}
\soulregister\citep7 
\soulregister\citet7 
\soulregister\citealp7 
\newsavebox{\measurebox} 
\usepackage{titlesec} 
\newcommand{\V}[1]{\boldsymbol{#1}} 

\journal{}

\makeatletter
\def\@author#1{\g@addto@macro\elsauthors{\normalsize%
    \def\baselinestretch{1}%
    \upshape\authorsep#1\unskip\textsuperscript{%
      \ifx\@fnmark\@empty\else\unskip\sep\@fnmark\let\sep=,\fi
      \ifx\@corref\@empty\else\unskip\sep\@corref\let\sep=,\fi
      }%
    \def\authorsep{\unskip,\space}%
    \global\let\@fnmark\@empty
    \global\let\@corref\@empty  
    \global\let\sep\@empty}%
    \@eadauthor={#1}
}
\makeatother

\titleformat{\paragraph}
{\normalfont\normalsize\itshape}{\theparagraph}{1em}{}
\titlespacing*{\paragraph}
{0pt}{3.25ex plus 1ex minus .2ex}{1.5ex plus .2ex}

\begin{document}

\begin{frontmatter}



\title{Non-local plasticity effects on notch fracture mechanics}


\author{Emilio Mart\'{\i}nez-Pa\~neda\corref{cor1}\fnref{CAM}}
\ead{mail@empaneda.com}

\author{Susana {del Busto}\fnref{Uniovi}}

\author{Covadonga Beteg\'on\fnref{Uniovi}}

\address[CAM]{Department of Engineering, Cambridge University, CB2 1PZ Cambridge, UK}

\address[Uniovi]{Department of Construction and Manufacturing Engineering, University of Oviedo, Gij\'on 33203, Spain}

\cortext[cor1]{Corresponding author. Tel: +44 1223 7 48525.}

\begin{abstract}
We investigate the influence of gradient-enhanced dislocation hardening on the mechanics of notch-induced failure. The role of geometrically necessary dislocations (GNDs) in enhancing cracking is assessed by means of a mechanism-based strain gradient plasticity theory. Both stationary and propagating cracks from notch-like defects are investigated through the finite element method. A cohesive zone formulation incorporating monotonic and cyclic damage contributions is employed to address both loading conditions. Computations are performed for a very wide range of length scale parameters and numerous geometries are addressed, covering the main types of notches. Results reveal a strong influence of the plastic strain gradients in all the scenarios considered. Transitional combinations of notch angle, radius and length scale parameter are identified that establish the regimes of GNDs-relevance, laying the foundations for the rational application of gradient plasticity models in damage assessment of notched components.
\end{abstract}


\end{frontmatter}



\section{Introduction}
\label{Sec:Introduction}

Heterogeneous plastic deformation requires additional dislocations to ensure geometric compatibility. These \emph{geometrically necessary dislocations} (GNDs) contribute mainly to material work hardening, rather than plastic straining, by acting as obstacles to the motion of \emph{statistically stored dislocations} (SSDs). Hence, the confinement of large gradients of plastic strain in a small volume translates into an increase of the strengthening and the hardening. This change in material response has been consistently observed in micron-scale tests (\emph{smaller is stronger}) such as indentation \cite{Nix1998}, bending \cite{Stolken1998} or torsion \cite{Fleck1994}, among many other. As a consequence, significant efforts have been devoted to the development of \emph{strain gradient} plasticity (SGP) theories, aiming to enrich conventional plasticity by incorporating the influence of GNDs (see \cite{Evans2009,IJSS2016} and references therein). While the investigation of gradient effects was initially motivated by growing interest in micro-technology, the influence of this size dependent plastic phenomenon extends beyond micron-scale applications, as plastic strains vary over microns in a wide range of engineering designs. GNDs have proven to have a significant effect on fracture \cite{Wei1997,Komaragiri2008}, fatigue \cite{Brinckmann2008,Brinckmann2008a}, strengthening on TRIP steels and fiber-reinforced materials \cite{Mazzoni-Leduc2008,Legarth2010}, hydrogen embrittlement \cite{IJHE2016,AM2016}, friction and contact \cite{Nielsen2016,Song2016}, void growth \cite{Niordson2007}, and damage \cite{Voyiadjis2006,Aifantis2016}. The role of GNDs on structural integrity assessment has attracted increasing attention in recent years; stress- and strain-based gradient theories have shown that GNDs near the crack tip promote local strain hardening and lead to a much higher stress level as compared with classic plasticity predictions \cite{Shishvan2016,CM2017}. Mart\'{\i}nez-Pa\~neda and co-workers \cite{IJSS2015,IJP2016} extended the analysis of crack tip fields to the finite deformation framework, showing that this stress elevation is substantially higher when large deformations are accounted for. Their parametric studies show that the physical length over which gradient effects prominently enhance crack tip stresses may span tens of $\mu$m, highlighting the need to incorporate this GND-effect in many damage models. However, modeling efforts have been restricted to cracked specimens and the influence of GNDs on the structural integrity assessment of notched components has not been addressed yet.\\

Many mechanical failures originate from notch-like defects and flaws accidentally introduced in service or during the manufacturing process. Numerous studies have been conducted to model the notch-induced rise in local stresses and subsequent cracking (see, e.g, the review by Ayatollahi et al. \cite{Ayatollahi2016}). The use of cohesive zone formulations has particularly gained popularity in this regard, as the cohesive traction-separation law constitutes a suitable tool to characterize cracking initiation and subsequent failure. Gomez and Elices used the cohesive zone model to develop a fracture criterion for both sharp and blunt V-notches \cite{Gomez2003,Gomez2004}, later extended to U-notches in linear elastic materials \cite{Gomez2006}. Olden et al. \cite{Olden2007} investigated hydrogen assisted cracking in notched samples through a hydrogen-dependent cohesive zone formulation. More recently, Cendon et al. \cite{Cendon2015} addressed fracture on coarse-grained polycrystalline graphite by means of an embedded cohesive crack technique \cite{Sancho2007}. Other popular approaches involve the use of Strain Energy Density criteria (see the contributions by Berto and Lazzarin \cite{Berto2009,Berto2014}).\\

In this work, strain gradient effects on notch-induced fracture are for the first time investigated. The role of GNDs in elevating the stresses ahead of notch-like defects and subsequently enhancing crack propagation is thoroughly examined under both monotonic and cyclic loading conditions. Crack tip stresses, critical loads and fatigue crack growth rates have been obtained over a wide range of length scales for different notch configurations. Finite element computations reveal important differences with conventional plasticity theory and unfold the relevance of non-local plasticity effects in notch mechanics. 

\section{Numerical framework}
\label{Sec:NumScheme}

The role of non-local plasticity effects in enhancing monotonic and cyclic damage ahead of notches is here investigated by means of a cohesive zone formulation and strain gradient plasticity. Section \ref{Sec:MSG} describes the adopted mechanism-based strain gradient (MSG) plasticity formulation and its numerical implementation. Section \ref{Sec:CZM} provides details of the cyclic-dependent cohesive zone formulation and presents different techniques employed to deal with the mechanical instabilities intrinsically associated with these models. Section \ref{Sec:FEmodel} outlines the boundary value problems under consideration and the finite element (FE) implementation.

\subsection{MSG plasticity}
\label{Sec:MSG}

\subsubsection{Constitutive prescriptions}

Grounded on the physical notion of GNDs, generated to accommodate lattice curvature due to non-uniform plastic deformation, SGP theories relate the yield strength (or the plastic work) to both strains and strain gradients; thereby introducing a length scale in the material description. At the phenomenological level, strain gradient models aim at capturing this gradient-enhanced dislocation hardening in poly-crystalline metals in an average sense, without explicitly accounting for the crystal lattice, nor for the behavior of internal grain boundaries. The length parameter is therefore generally obtained by fitting experimental measurements of micro-tests through a specific SGP theory (in a way that resembles the fitting of the strain hardening exponent by means of a specific power law). Both mechanism-based \cite{Gao1999,Huang2004} and phenomenological \cite{Fleck2001,Gudmundson2004} isotropic SGP constitutive laws have been proposed - we here focus on the former.\\

The mechanism-based theory of strain gradient plasticity was proposed by Gao and co-workers \cite{Gao1999,Huang2000} based on a multiscale framework linking the microscale concept of SSDs and GNDs to the mesoscale notion of plastic strains and strain gradients. Unlike other SGP formulations, MSG plasticity introduces a linear dependence of the square of plastic flow stress on strain gradient. This linear dependence was largely motivated by the nano-indentation experiments of Nix and Gao \cite{Nix1998} and comes out naturally from Taylor's dislocation model \cite{Taylor1938}, on which MSG plasticity is built. Therefore, while all continuum formulations have a strong phenomenological component, MSG plasticity differs from all existing phenomenological theories in its mechanism-based guiding principles. The constitutive equations common to mechanism-based theories are summarized below; more details can be found in the original works \cite{Gao1999,Huang2000}.\\

In MSG plasticity, since the Taylor model is adopted as a founding principle, the shear flow stress $\tau$ is formulated in terms of the total dislocation density $\rho$ as
\begin{equation} \label{eq:1MSG}
\tau = \alpha \mu b \sqrt{\rho}
\end{equation}
Here, $\mu$ is the shear modulus, $b$ is the magnitude of the Burgers vector and $\alpha$ is an empirical coefficient that is generally taken to be 0.5. The dislocation density is composed of the sum of the density $\rho_S$ for SSDs and the density $\rho_G$ for GNDs as
\begin{equation} \label{Eq:2MSG}
\rho = \rho_S + \rho_G
\end{equation}
The GND density $\rho_G$ is related to the effective plastic strain gradient $\eta^{p}$ by: 
\begin{equation} \label{Eq:3MSG}
\rho_G = \overline{r}\frac{\eta^{p}}{b}
\end{equation}
\noindent where $\overline{r}$ is the Nye-factor which is assumed to be 1.90 for face-centered-cubic (fcc) polycrystals. Following Fleck and Hutchinson \cite{Fleck1997}, Gao et al. \cite{Gao1999} used three quadratic invariants of the plastic strain gradient tensor to represent the effective plastic strain gradient $\eta^{p}$ as
\begin{equation}
\eta^{p}=\sqrt{c_1 \eta^{p}_{iik} \eta^{p}_{jjk} + c_2 \eta^{p}_{ijk} \eta^{p}_{ijk} + c_3 \eta^{p}_{ijk} \eta^{p}_{kji}}
\end{equation}

The coefficients were determined to be equal to $c_1=0$, $c_2=1/4$ and $c_3=0$ from three dislocation models for bending, torsion and void growth, leading to
\begin{equation}
\eta^{p}=\sqrt{\frac{1}{4} \V{\eta^{p}} \cdot \V{\eta^{p}}}
\end{equation}

\noindent where the components of the strain gradient tensor are obtained by,
\begin{equation}
\eta^{p}_{ijk}= \varepsilon^{p}_{ik,j}+\varepsilon^{p}_{jk,i}-\varepsilon^{p}_{ij,k}
\end{equation}

The tensile flow stress $\sigma_{flow}$ is related to the shear flow stress $\tau$ by,
\begin{equation} \label{eq:4MSG}
\sigma_{flow} =M\tau
\end{equation}
\noindent where $M$ is the Taylor factor, taken to be 3.06 for fcc metals. Rearranging Eqs. (\ref{eq:1MSG}-\ref{Eq:3MSG}) and Eq. (\ref{eq:4MSG}) yields
\begin{equation} \label{Eq5MSG}
\sigma_{flow} =M\alpha \mu b \sqrt{\rho_{S}+\overline{r}\frac{\eta^{p}}{b}}
\end{equation}
The SSD density $\rho_{S}$ can be determined from (\ref{Eq5MSG}) knowing the relation in uniaxial tension between the flow stress and the material stress-strain curve as follows
\begin{equation} \label{Eq6MSG}
\rho_{S} = [\sigma_{ref}f(\varepsilon^{p})/(M\alpha \mu b)]^2
\end{equation}
Here $\sigma_{ref}$ is a reference stress and $f$ is a non-dimensional function of the plastic strain $\varepsilon^{p}$ determined from the uniaxial stress-strain curve. Substituting back into (\ref{Eq5MSG}), $\sigma_{flow}$ yields
\begin{equation} \label{Eq:Sflow}
\sigma_{flow} =\sigma_{ref} \sqrt{f^2(\varepsilon^{p})+\ell \eta^{p}}
\end{equation}
\noindent where $\ell$ is the intrinsic material length. Hence, gradient effects become negligible and the flow stress recovers the conventional plasticity solution if the characteristic length of plastic deformation outweighs the GNDs-related term $\ell \eta^{p}$. 

\subsubsection{Numerical implementation}

The conventional theory of mechanism-based strain gradient (CMSG) plasticity \cite{Huang2004} is here chosen since, unlike its higher order counterpart, it does not suffer convergence problems in finite strain fracture problems \cite{Hwang2003,IJSS2015}. As discussed in \cite{Qu2004}, the Taylor dislocation model gives the flow stress dependent on both the equivalent plastic strain $\varepsilon^p$ and effective plastic strain gradient $\eta^p$
\begin{equation}
\dot{\sigma}=\frac{\partial \sigma}{\partial \varepsilon^p} \dot{\varepsilon}^p + \frac{\partial \sigma}{\partial \eta^p} \dot{\eta}^p
\end{equation}

\noindent such that, for a plastic strain rate $\dot{\V{\varepsilon}}^p$ proportional to the deviatoric stress $\V{\sigma}'$, a self contained constitutive model cannot be obtained due to $\dot{\eta}^p$. In order to overcome this situation without employing higher order stresses, Huang et al. \cite{Huang2004} adopted a viscoplastic formulation to obtain $\dot{\varepsilon}^p$ in terms of the effective stress $\sigma_e$ rather than its rate $\dot{\sigma}_e$
\begin{equation}
\dot{\varepsilon}^{p} = \dot{\varepsilon} \left [\frac{\sigma_e}{\sigma_{ref} \sqrt{f^{2}(\varepsilon^{p})+\ell \eta^{p}}} \right]^{m}
\end{equation}

\noindent where the rate-independent limit is achieved by replacing the reference strain with the effective strain rate $\dot{\varepsilon}$ and taking the exponent to fairly large values ($m\geq20$) \cite{Huang2004}. The governing equations are therefore essentially the same as those in conventional plasticity and the plastic strain gradient comes into play through the incremental plastic modulus; the constitutive equation is given by,
\begin{equation}
\dot{\V{\sigma}}=K \textnormal{tr} \left( \dot{ \V{\varepsilon}} \right) \V{\delta}+2\mu \left\{ \dot{\V{\varepsilon}}' - \frac{3\dot{\varepsilon}}{2\sigma_e}\left[\frac{\sigma_e}{\sigma_{ref} \sqrt{f^{2}(\varepsilon^{p})+\ell \eta^{p}}} \right]^{m} \dot{\V{\sigma}}' \right\}
\end{equation}

\noindent Here $K$ being the bulk modulus and $\V{\delta}$ the Kronecker delta. Further, $\V{\sigma}$ is the Cauchy stress tensor and the work-conjugate strain tensor is denoted by $\V{\varepsilon}$. Since higher order terms are not involved, the FE implementation is relatively straightforward. The plastic strain gradient is obtained by numerical differentiation within the element: the plastic strain increment is interpolated through its values at the Gauss points in the isoparametric space and afterwards the increment in the plastic strain gradient is calculated by differentiation of the shape functions. In the present finite strain analysis, rigid body rotations for the strains and stresses are carried out by means of the Hughes and Winget's algorithm \cite{Hughes1980} and the strain gradient is obtained from the deformed configuration (see \cite{IJSS2015}). Although higher order terms are required to model effects of dislocation blockage at impermeable boundaries, one should note that higher order boundary conditions have essentially no effect on the stress distribution at a distance of more than 10 nm away from the crack tip in MSG plasticity \citep{Shi2001,Qu2004}, well below its lower limit of physical validity - the model represents an average of dislocation activities and it is therefore only applicable at a scale much larger than the average dislocation spacing ($\approx$ 100 nm). 

\subsection{Cohesive zone model}
\label{Sec:CZM}

We model cracking ahead of the notch-tip under monotonic and periodic loading by means of a potential-based cohesive zone formulation. In the interest of brevity, the description of the traction-separation relation and its numerical implementation are particularized for the  conditions under consideration: pure mode I problems where the cohesive interface lies on the symmetry line. For details on the implementation of cohesive elements within a conventional finite element framework the reader is referred to \cite{EFM2017}.

\subsubsection{Constitutive traction-separation law}
\label{Sec:TractionSeparationLaw}

The pivotal ingredient of cohesive zone models is the traction-separation law that governs material degradation and separation. The exponentially decaying cohesive law proposed by Xu and Needleman \cite{Xu1999} is here adopted. The cohesive response is therefore characterized by the relation between the normal traction $T_n$ and the corresponding displacement jump $\Delta_n$ as,
\begin{equation}
 T_n= \frac{\phi_n}{\delta_n} \exp \left( - \frac{\Delta_n}{\delta_n} \right) \frac{\Delta_n}{\delta_n}
\end{equation}

\noindent where $\phi_n$ denotes the normal work of separation, which is given by,
\begin{equation}
\phi_n= \exp(1) \sigma_{max,0} \delta_n
\end{equation} 

\begin{figure}[H]
\centering
\includegraphics[scale=0.8]{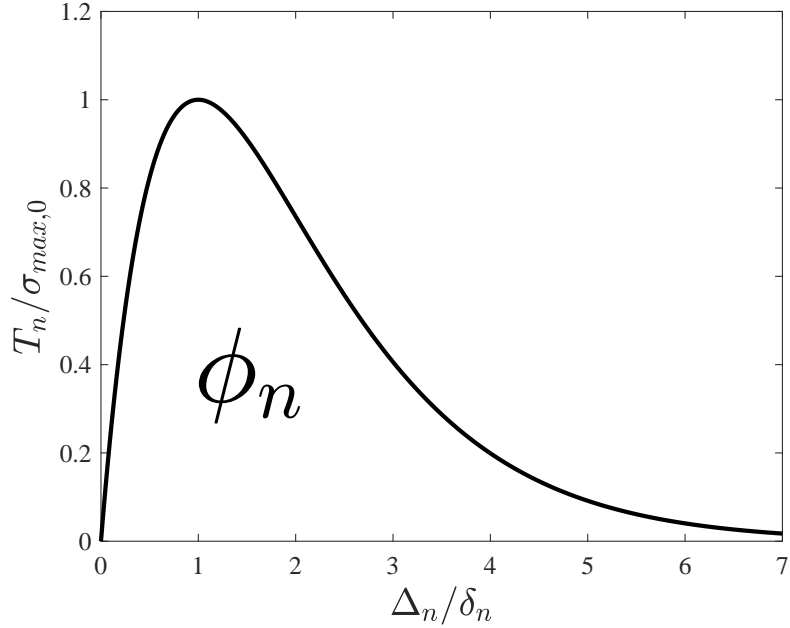}
\caption{Traction-separation law characterizing the cohesive zone model in the absence of cyclic damage degradation.}
\label{fig:CoheLaw}
\end{figure}

\noindent Such that, grounded on atomistic calculations \cite{Xu1999}, the normal response is assumed to follow an exponential form, as depicted in Fig. \ref{fig:CoheLaw}. Here, $\sigma_{max}$ is the interface normal strength, while $\delta_n$ refers to the characteristic opening length in the normal direction. The subscript 0 indicates that $\sigma_{max,0}$ is the \emph{initial} normal strength, which can be reduced due to, e.g., fatigue or environmental damage \cite{EFM2017}. For a given shape of the traction-separation curve, the cohesive response can be fully characterized by two parameters, the cohesive energy $\phi_n$ and the critical cohesive strength $\sigma_{max,0}$.

Cyclic damage is incorporated by means of the irreversible cohesive zone model proposed by Roe and Siegmund \cite{Roe2003}. The model incorporates (i) loading-unloading conditions, (ii) accumulation of damage during subcritical cyclic loading, and (iii) crack surface contact. A damage mechanics approach is adopted to capture the cohesive properties degradation as a function of the number of cycles. An effective cohesive zone traction is consequently defined as,
\begin{equation}
\tilde{\V{T}}=\frac{\V{T}}{(1-D)}
\end{equation}

\noindent with $D$ being a damage variable that represents the effective surface density of micro defects in the interface. Accordingly, the current or effective cohesive strength $\sigma_{max}$ is related to the initial cohesive strength $\sigma_{max,0}$ as,
\begin{equation}
\sigma_{max}=\sigma_{max,0} (1 - D)
\end{equation}

\begin{figure}[H]
\centering
\includegraphics[scale=0.9]{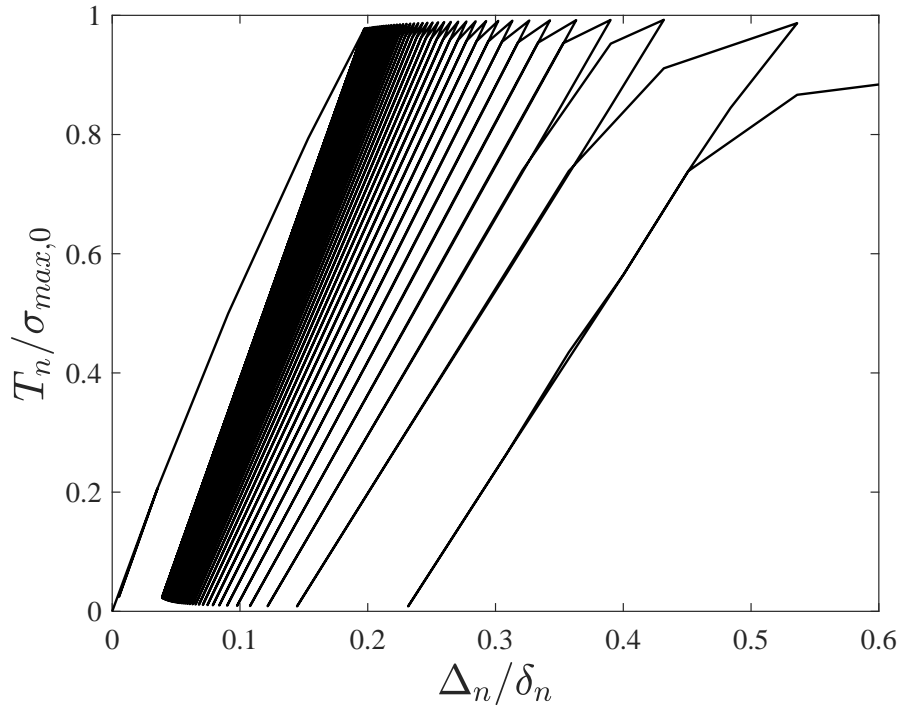}
\caption{Representative cohesive response under stress-controlled cyclic loading conditions.}
\label{fig:CoheLawF}
\end{figure}

A damage evolution law is defined that incorporates the relevant features of continuum damage approaches, namely: (i) damage accumulation starts if a deformation measure is greater than a critical magnitude, (ii) the increment of damage is related to the increment of deformation, and (iii) an endurance limit exists bellow which cyclic loading can proceed infinitely without failure. From these considerations, cyclic damage evolution is given by,
\begin{equation}\label{Eq:Damage}
\dot{D}_c= \frac{|\dot{\Delta}_n|}{\delta_{\Sigma}} \left[ \frac{T_n}{\sigma_{max}} - \frac{\sigma_f}{\sigma_{max,0}} \right] H \left( \bar{\Delta}_n -  \delta_n \right)
\end{equation}

\noindent with $\bar{\Delta}_n=\int |\dot{\Delta}_n| dt$ and $H$ denoting the Heaviside function. Two new parameters have been introduced: $\sigma_f$, the cohesive endurance limit and $\delta_{\Sigma}$, the accumulated cohesive length. The latter is used to scale the normalized increment of the effective material separation. The model must also incorporate damage due to monotonic loading; as a consequence, the damage state is defined as the maximum of the cyclic and monotonic contributions,
\begin{equation}
D= \int \textnormal{max} \left( \dot{D}_c, \dot{D}_m \right) dt
\end{equation}

\noindent being $\dot{D}_m$ generally defined as,
\begin{equation}
\dot{D}_m = \frac{ \left. \textnormal{max} \left( \Delta_n \right) \right|_{t_i} -  \left. \textnormal{max} \left( \Delta_n \right) \right|_{t_{i-1}}}{4 \delta_n}
\end{equation}

\noindent and updated only when the largest stored value of $\Delta_n$ is greater than $\delta_N$. Here, $t_{i-1}$ denotes the previous time increment and $t_i$ the current one. The same modeling framework can be therefore employed for monotonic and cyclic loading case studies, as it is the case of the present work. Moreover, the cohesive response must be defined for the cases of unloading/reloading, compression, and contact between the crack faces. Unloading is defined based on the analogy with an elastic-plastic material undergoing damage. Thereby, unloading takes place with the stiffness of the cohesive zone at zero separation, such that
\begin{equation}
T_n=T_{max} + \left( \frac{\exp(1) \sigma_{max}}{\delta_n} \right) \left( \Delta_n - \Delta_{max} \right)
\end{equation}

\noindent where $\Delta_{max}$ is maximum separation value that has been attained and $T_{max}$ its associated traction quantity. Compression behavior applies when the unloading path reaches $\Delta_n=0$ at $T_n < 0$. In such circumstances, the cohesive response is given by,
\begin{align}
T_n = & \frac{\phi_n}{\delta_n} \left( \frac{\Delta_n}{\delta_n} \right) \exp \left( - \frac{\Delta_n}{\delta_n} \right) + T_{max} - \sigma_{max} \exp(1) \frac{\Delta_{max}}{\delta_n} \nonumber \\
& + \alpha \sigma_{max,0} \exp(1) \frac{\Delta_n}{\delta_n} \exp \left( - \frac{\Delta_n}{\delta_n} \right)
\end{align}

\noindent being $\alpha$ a penalty factor that is taken to be equal to 10, following \cite{Roe2003}. Contact conditions are enforced if $\Delta_n$ is negative and the cohesive element has failed completely ($D=1$). At this instance the cohesive law renders,
\begin{equation}
T_n= \alpha \sigma_{max,0} \exp(1) \exp \left( - \frac{\Delta_n}{\delta_n} \right) \frac{\Delta_n}{\delta_n} 
\end{equation}

\noindent where friction effects have been neglected. Fig. \ref{fig:CoheLawF} shows the cohesive response obtained under stress-controlled cyclic loading $\Delta \sigma / \sigma_{max,0}=1$ with a zero stress ratio. The accumulated separation increases with the number of loading cycles, such that it becomes larger than $\delta_n$ and fatigue damage starts to play a role, lowering the stiffness and the cohesive strength.

\subsubsection{Control algorithm}
\label{Sec:ControlAlgorithm}

The softening part of the traction-separation law gives rises to a local stiffness degradation in the corresponding cohesive elements, which triggers elastic snap-back instabilities. Hence, at the point where the stress reaches the peak strength of the interface, quasi-static finite element computations are unable to converge to an equilibrium solution, hindering the modeling of the post-instability behavior. We here propose prescribing the opening displacement at the incipient crack while leaving the remote loading as a variable. This can be achieved by means of mixed FE-Rayleigh-Ritz formulations \cite{Tvergaard1976} or control algorithms \cite{Segurado2004,JMPS2017}; the latter approach is here adopted. Hence, as first described by Segurado and Llorca \cite{Segurado2004}, the simultaneous reduction of the load and the displacement at the load point can be captured by finding a variable that increases monotonically during the whole loading history. Here, in the context of a symmetric model, we choose to prescribe the average opening displacement ahead of the notch tip. An auxiliary element is created that connects the vertical displacement of the nodes ahead of the notch tip ($N_1, N_2, \dots , N_n$) with a control node $N_c$,
\begin{equation}
\begin{pmatrix}   0 & 0 & \cdots & 0 \\
  0 & 0 & \cdots & 0 \\
  \vdots  & \vdots  & \ddots & \vdots  \\
  1 & 1 & \cdots & 0  \end{pmatrix} \begin{pmatrix} u_y^{N_1} \\ u_y^{N_2}\\ \vdots \\ u_y^{N_c} \end{pmatrix}= \begin{pmatrix} f_y^{N_1} \\ f_y^{N_2}\\ \vdots \\ f_y^{N_c} \end{pmatrix}
\end{equation}

\noindent such that the average opening displacement is linearly related to the vertical force in the control node. The displacement in such control node is then equated to the vertical load in one of the nodes in the outer boundary $N_L$, where a remote displacement is generally prescribed. A second auxiliary element is defined for this purpose,
\begin{equation}
\begin{pmatrix}   0 & 1 \\
  0 & 0  \end{pmatrix} \begin{pmatrix} u_y^{N_L} \\ u_y^{N_c} \end{pmatrix}= \begin{pmatrix} f_y^{N_L} \\ f_y^{N_c} \end{pmatrix}
\end{equation}

In that way the average opening is prescribed by imposing a vertical force on the control node, and the displacement at the outer boundary is an outcome of the equilibrium solution.\\

The aforementioned control algorithm cannot, however, be used for cyclic loading, where we want to make sure that the external load follows a specific (sinusoidal) behavior. In some of the fatigue computations numerical convergence has been facilitated by employing the viscous regularization technique proposed by Gao and Bower \cite{Gao2004}. Such scheme leads to accurate results if the viscosity coefficient is sufficiently small - a sensitivity study has been conducted in the few cases where viscous regularization was needed.

\subsection{Finite element implementation}
\label{Sec:FEmodel}

The aforementioned numerical model is implemented in the commercial finite element package ABAQUS. The MSG plasticity model is incorporated by means of a user material subroutine (UMAT), while a user element subroutine (UEL) is employed for the cohesive element formulation. Results post-processing is carried out in MATLAB by making use of \emph{Abaqus2Matlab} \cite{AES2017}, a novel tool that connects the two well-known aforementioned software suites.\\

\begin{figure}[H]
\makebox[\linewidth][c]{%
        \begin{subfigure}[b]{0.4\textwidth}
                \centering
                \includegraphics[scale=1]{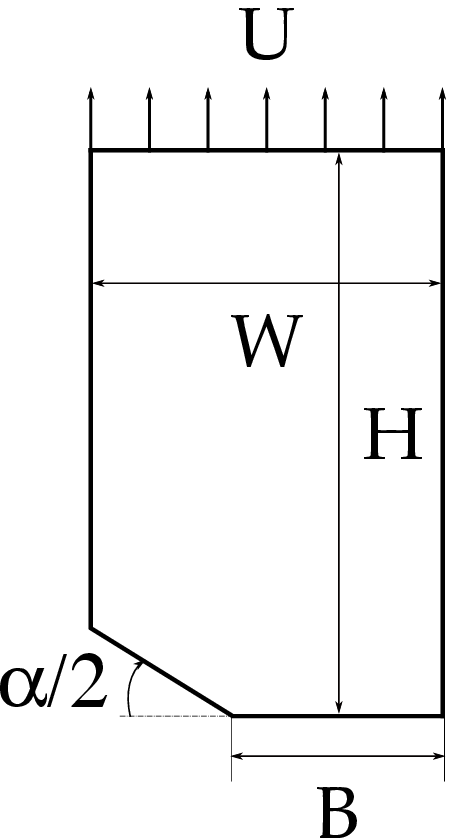}
                \caption{}
                \label{fig:Micromachining}
        \end{subfigure}
        \begin{subfigure}[b]{0.42\textwidth}
                \raggedleft
                \includegraphics[scale=1]{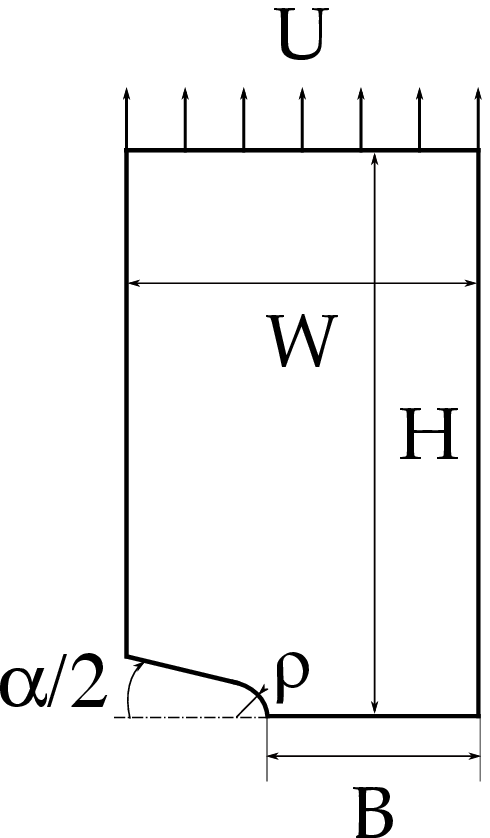}
                \caption{}
                \label{fig:Laser}
        \end{subfigure}
        \begin{subfigure}[b]{0.4\textwidth}
                \raggedleft
                \includegraphics[scale=1]{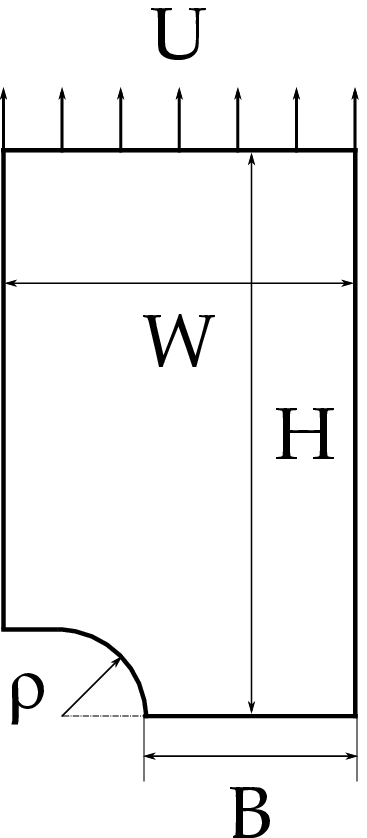}
                \caption{}
                \label{fig:Laser}
        \end{subfigure}        
        }
       
        \caption{Geometry of the notched plates under consideration, (a) sharp V-notch, (b) blunted V-notch, and (c) U-notch.}\label{fig:Notches}
\end{figure}

We illustrate the effect of strain gradient theories on notch mechanics by investigating the main types of notches. Namely, (i) sharp V-notches with different angles, (ii) blunted V-notches with different tip radii, and (iii) U-notches with different radii. Hence, as described in Fig. \ref{fig:Notches}, a notched plate of height $H=80$ mm, width $W=0.3125H$, and notch ligament $B=0.25H$, is considered as reference geometry in all cases. Only the upper half of the specimen is shown, as we take advantage of symmetry. Plane strain conditions are assumed and all computations account for large strains and rotations. After a sensitivity study, a very fine mesh is used, with the size of the elements ahead of the crack being significantly smaller than the characteristic length of the fracture process zone ($\approx R_0/500$),
\begin{equation}\label{Eq:R0}
R_0 = \frac{1}{3 \pi \left( 1 - \nu^2 \right)} \frac{E \phi_n}{\sigma_Y^2} = \frac{1}{3 \pi} \left( \frac{K_0}{\sigma_Y} \right)^2
\end{equation}

\noindent Here, $E$ is Young's modulus, $\sigma_Y$ the yield stress and $\nu$ Poisson's ratio. Higher order elements are used in all cases: 8-node quadrilateral elements with reduced integration are employed to model the bulk response and crack initiation and growth are captured by 6-node quadrilateral cohesive elements with 12 integration points. A reference stress intensity factor is defined from the cohesive crack,
\begin{equation}
K_0=\sqrt{\frac{E \phi_n}{1 - \nu^2}}
\end{equation}

\noindent such that an associated reference remote stress, $\sigma_0$, can be defined from fracture mechanics considerations ($K=\sigma \sqrt{\pi a}$, for a geometrical factor equal to 1). Accordingly, one can make use of a reference external load, $P_0$, by dividing the reference remote stress by the notch ligament and the thickness. Dimensional analysis of this set of parameters reveals that the solution, given by the external force $P$, depends on the following dimensionless combinations,
\begin{equation}
\frac{P}{P_0}= F \left(\frac{\rho}{R_0}, \, \frac{\sigma_Y}{E}, \, \frac{\Delta a}{R_0}, \, \frac{\sigma_{max}}{\sigma_Y}, \, \frac{\ell}{R_0}, \,  n, \, \nu, \, \alpha \right) 
\end{equation}

\noindent where $\rho$ denotes the notch radius (see Fig. \ref{fig:Notches}), $\Delta a$ the crack extension and $F$ is a dimensionless function of the arguments displayed.
We investigate the notch fracture resistance of a steel of $\sigma_Y/E=0.003$, Poisson's ratio $\nu=0.3$ and an isotropic hardening response given by,
\begin{equation}
\sigma = \sigma_Y \left(1 + \frac{E \varepsilon_p}{\sigma_Y} \right)^{\left(1/n\right)}
\end{equation}

\noindent with the strain hardening exponent being equal to $n=5$. The reference stress in Eq. (\ref{Eq:Sflow}) is therefore given by $\sigma_{ref}=\sigma_Y \left( E / \sigma_Y \right)^{(1/n)}$ and $f(\varepsilon^p)=\left(\varepsilon^p + \sigma_Y/E \right)^{(1/n)}$. The length scale parameter is varied over a very wide range so as to cover the whole spectrum of experimentally reported values.

\section{Results}
\label{Sec:Results}

The role of geometrically necessary dislocations in compromising the structural integrity of notched components is here investigated by strain gradient plasticity computations of: (i) stationary notch tip fields (Section \ref{Sec:Stationary}), (ii) cohesive crack propagation under monotonic loading conditions (Section \ref{Sec:Monotonic}), and (iii) fatigue crack growth (Section \ref{Sec:Cyclic}). 

\subsection{Stationary notch tip fields}
\label{Sec:Stationary}

We first investigate the influence of plastic strain gradients ahead of the notch tip in elevating the stresses so as to isolate the analysis of gradient effects from the cohesive description of damage. Hence, the opening stress $\sigma_{22}$ is computed for the three different geometries outlined in Fig. \ref{fig:Notches}, considering for each case different notch radii and angles. Results are presented ahead of the notch with the distance to the tip normalized by the reference size of the fracture process zone, given by the last expression in Eq. (\ref{Eq:R0}). Here, the reference stress intensity factor $K_0$ is taken as the external load $K_I$, which is computed from the stress at the remote boundary $\sigma_R$ (see description in Section \ref{Sec:FEmodel}). By considering a geometrical factor equal to 1 in all configurations, $R_0$ provides a quantitative description of the external load.\\

\begin{figure}[H]
\centering
\includegraphics[scale=1.05]{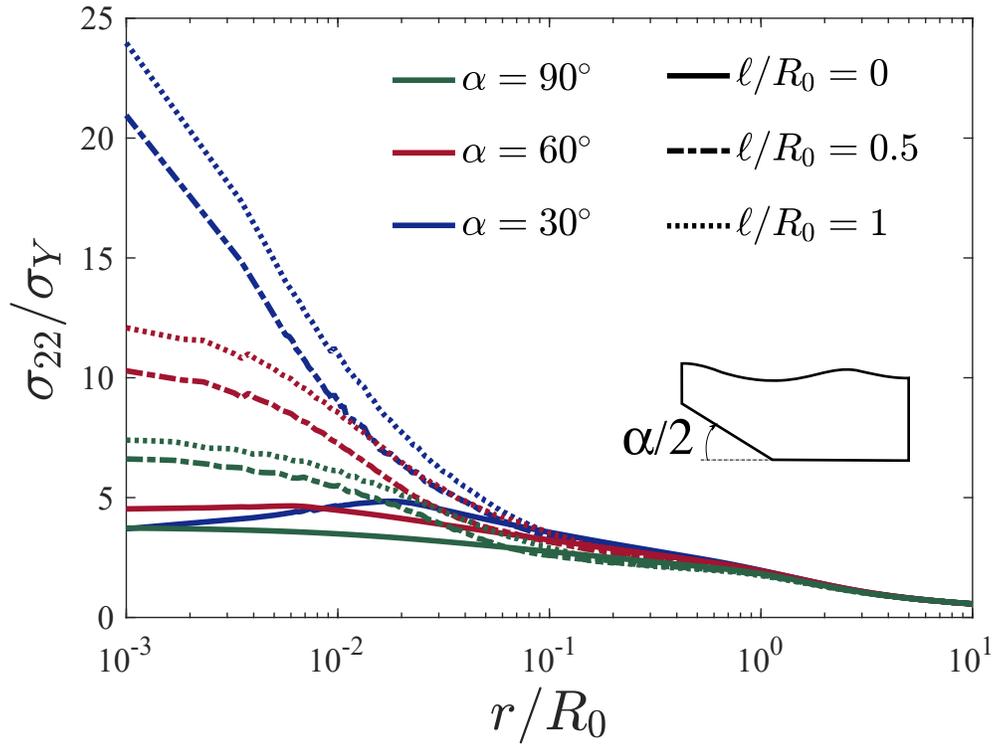}
\caption{Notch tip opening stresses for the sharp V-notch case. Results are shown for different angles and different values of the length scale parameter. Material properties: $\sigma_Y/E=0.003$, $\nu=0.3$, and $N=0.2$.}
\label{fig:CrackTipV}
\end{figure}

Finite element results obtained for the sharp V-notch geometry are shown in Fig. \ref{fig:CrackTipV}. The opening stress ahead of the extended notch plane is shown for three different angles of the initial notch opening ($\alpha=30^{\circ}$, $\alpha=60^{\circ}$ and $\alpha=90^{\circ}$) and three values of the length scale parameter ($\ell/R_0=0$, $\ell/R_0=0.5$ and $\ell/R_0=1$). The figure shows a very significant stress elevation, relative to the conventional plasticity case ($\ell/R_0=0$), when the GND-effect is accounted for. This outcome of the GND promoted hardening increases with the length parameter, in agreement with expectations, and is particularly enhanced, for a given external load, by decreasing the notch angle. The differences are particularly meaningful for the smallest radius, where the gradient-enhanced stresses are 4-5 times larger than the conventional plasticity predictions in the vicinity of the notch tip. This stress elevation, that falls short of attaining the theoretical lattice strength ($E/10$), is relevant in a domain that spans one-tenth of the plastic zone size ($R_0$ resembles the plastic zone length for this crack-like geometry); far from the notch tip both conventional and strain gradient plasticity solutions agree.

\begin{figure}[H]
\centering
\includegraphics[scale=1.1]{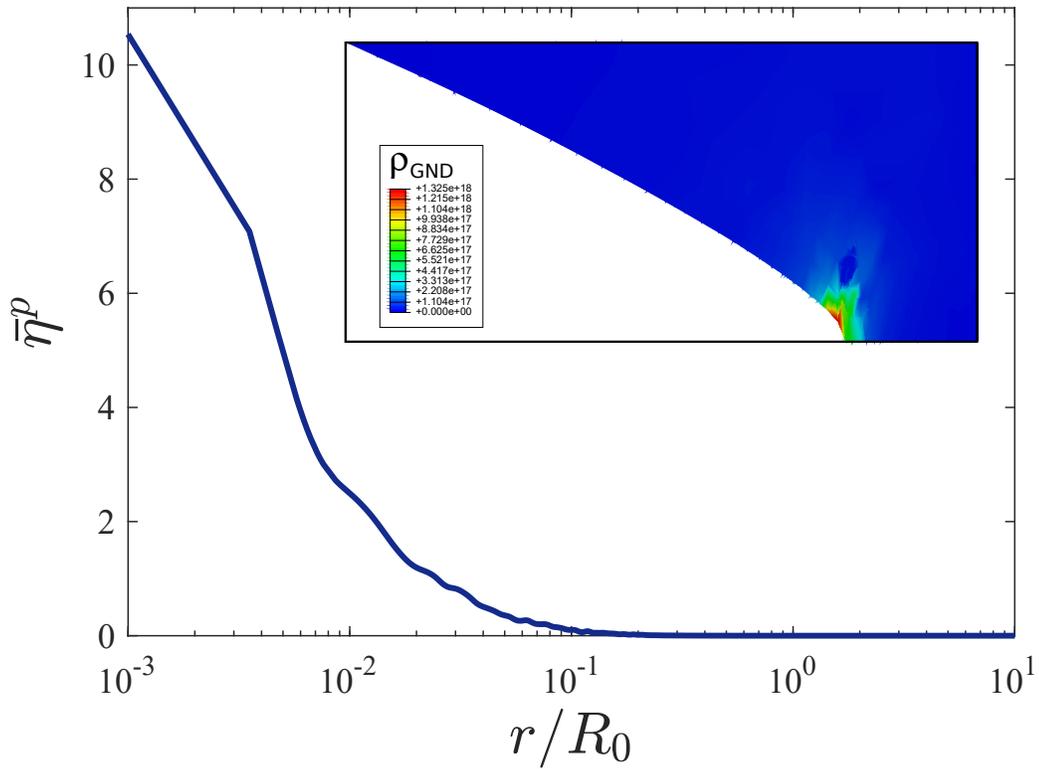}
\caption{Normalized effective plastic strain gradient $\bar{\eta}^p=R_0 \eta^p$ ahead of the notch tip for the sharp V-notch specimen with $\alpha=30^{\circ}$. The embedded figure represents the GND density contours in m$^{-2}$. Material properties: $\sigma_Y/E=0.003$, $\nu=0.3$, $N=0.2$, and $\ell/R_0=1$.}
\label{fig:Grad}
\end{figure}

Further insight is gained by looking at the effective plastic strain gradient ahead of the notch, along with the associated GND contours. Fig. \ref{fig:Grad} shows a normalized effective plastic strain gradient $\bar{\eta}^p=R_0 \eta^p$ for the geometry with a notch radius of $\alpha=30^{\circ}$ and a length scale parameter of $\ell/R_0=1$. Results reveal a very meaningful increase of the plastic strain gradients within a fraction of the plastic zone. GNDs are generated to accommodate this nonhomogeneous plastic deformation, promoting strain hardening and leading to notch tip stresses that are much larger than those predicted using conventional continuum theories.

\begin{figure}[H]
\centering
\includegraphics[scale=1.05]{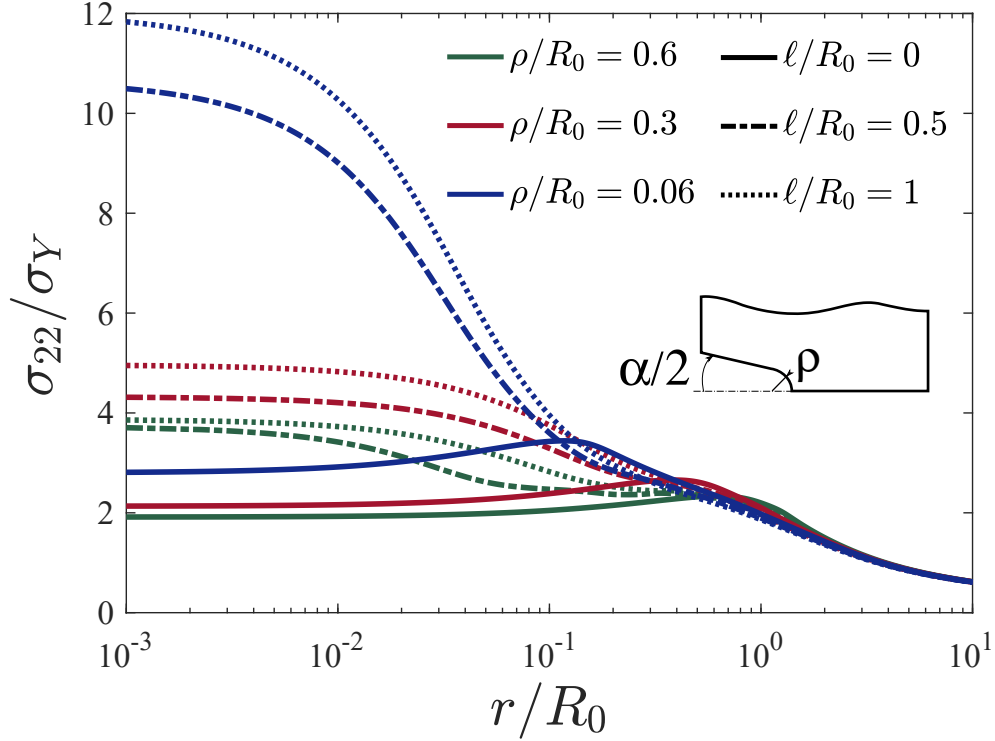}
\caption{Notch tip opening stresses for the blunted V-notch case. Results are shown for different notch radii and different values of the length scale parameter. Material properties: $\sigma_Y/E=0.003$, $\nu=0.3$, and $N=0.2$.}
\label{fig:CrackTipVb}
\end{figure}

The opening stress distribution is also computed for the blunted V-notch case and the results obtained are shown in Fig. \ref{fig:CrackTipVb}. Different notch radii have been considered and a notch angle of $\alpha=30^{\circ}$ has been chosen for all calculations related to the \emph{blunted} V-notch geometry in this work. Results reveal an increase of the stress level with decreasing the notch radius, as it could be expected. Again, both gradient-enhanced and conventional plasticity predictions agree far from the notch but differences arise as the distance to the tip decreases. The GND-enriched prediction leads to stresses close to the notch tip that are at least 2 times those of conventional plasticity, and that could be up to 4 times for the smallest notch radius considered.

\begin{figure}[H]
\centering
\includegraphics[scale=1.05]{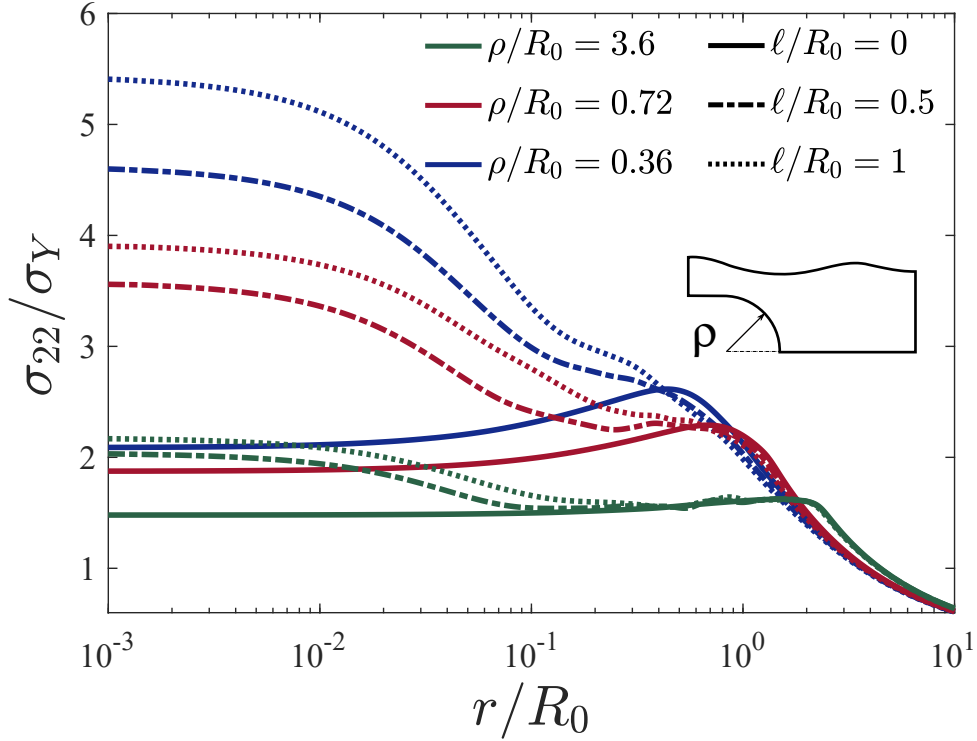}
\caption{Notch tip opening stresses for the U-notch case. Results are shown for different notch radii and different values of the length scale parameter. Material properties: $\sigma_Y/E=0.003$, $\nu=0.3$, and $N=0.2$.}
\label{fig:CrackTipU}
\end{figure}

Similar qualitative results are observed for the U-notch geometry (see Fig. \ref{fig:CrackTipU}). For a given load, the stresses increase with diminishing notch radius and significant differences between conventional and strain gradient plasticity solutions can be observed. Crack tip stresses are 1.5-2.5 times larger when GNDs are not neglected and the domain where these differences takes place can be on the order of $R_0$. This gradient-dominated region decreases significantly as the notch radius increases.

By comparing the results obtained for the three geometries under consideration one can see that the degree of stress elevation is higher for the crack-like sharp V-notch, as it could be expected \emph{a priori}. The differences in the peak stress level with conventional plasticity are higher for the blunted V-notch than for the U-notch, as the notch radii are smaller; the tip radius of a blunted V-notch is typically much smaller than the defect radius of a U-notched specimen. However, the inverse trend is obtained with respect to the size of the gradient dominated zone under the same external load. U-notch specimens show the largest physical length-scale over which strain gradients are prominent, followed by the blunted V-notch specimens. Smaller notch angles and radii lead to shorter GND domains (as they scale with the plastic zone region) but to much steeper gradients of plastic strain. Hence, the size of the defect characterizes the GND influence, which is bounded between two cases: (a) a micron-scale GND region with much higher stresses than those attained with conventional models, and (b) a larger gradient-dominated length with a lesser stress elevation.

\subsection{Monotonic loading}
\label{Sec:Monotonic}

\begin{figure}[H]
\centering
\includegraphics[scale=1.05]{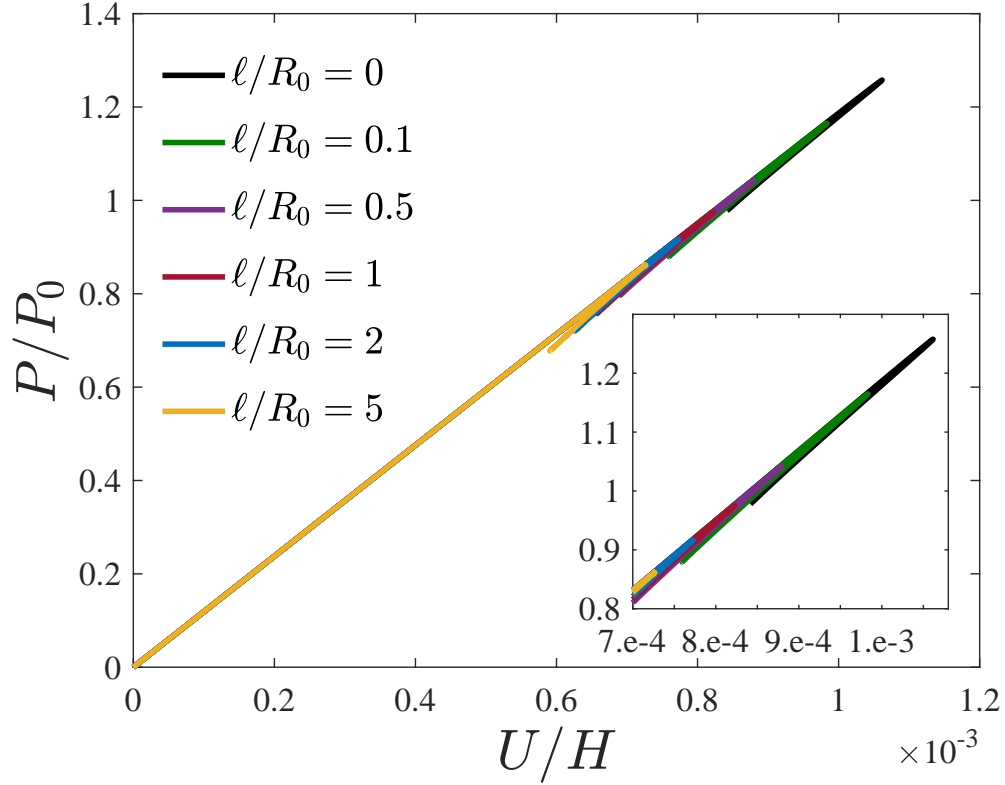}
\caption{Applied load versus remote strain for the sharp V-notch case with $\alpha=60^{\circ}$. Results are shown for both conventional plasticity and MSG plasticity with different values of the length scale parameter. Material properties: $\sigma_Y/E=0.003$, $\nu=0.3$, $N=0.2$ and $\sigma_{max,0}=3.5 \sigma_Y$.}
\label{fig:LoadDisp}
\end{figure}

Crack initiation and consequent propagation under monotonic loading conditions is subsequently investigated by making use of the cohesive zone formulation described in Section \ref{Sec:CZM}. The specimens are loaded by using a control algorithm (see Section \ref{Sec:ControlAlgorithm}) and the macroscopic response is captured beyond the point of unstable crack propagation. Fig. \ref{fig:LoadDisp} shows the force versus displacement curve obtained for the sharp V-notch specimen for the intermediate case of $\alpha=60^{\circ}$. Results are shown normalized, representing the abscissa axis a measure of the applied deformation. Both conventional plasticity and strain gradient plasticity have been considered, the latter through a wide range of length scale parameters.

As shown in Fig. \ref{fig:LoadDisp} the load increases up to a critical point, where a sudden snap-back response is observed as a consequence of the propagation of the crack from the notch tip. The use of the control algorithm described in Section \ref{Sec:ControlAlgorithm} enables to track the equilibrium solution throughout this unstable behavior where both the load and the displacement decrease. This critical point corresponds to the maximum load carrying capacity of the structure and will be subsequently denoted as $P_{max}$. Gradient-enriched results show how the stress elevation associated with dislocation hardening reduces $P_{max}$; more than a 30\% decrease in the maximum carrying capacity is observed for the largest value of $\ell$. The detrimental effect of GNDs on structural integrity is therefore not only restricted to infinitesimally sharp cracks but also present in notch-like defects.\\

\begin{figure}[H]
\centering
\includegraphics[scale=1.0]{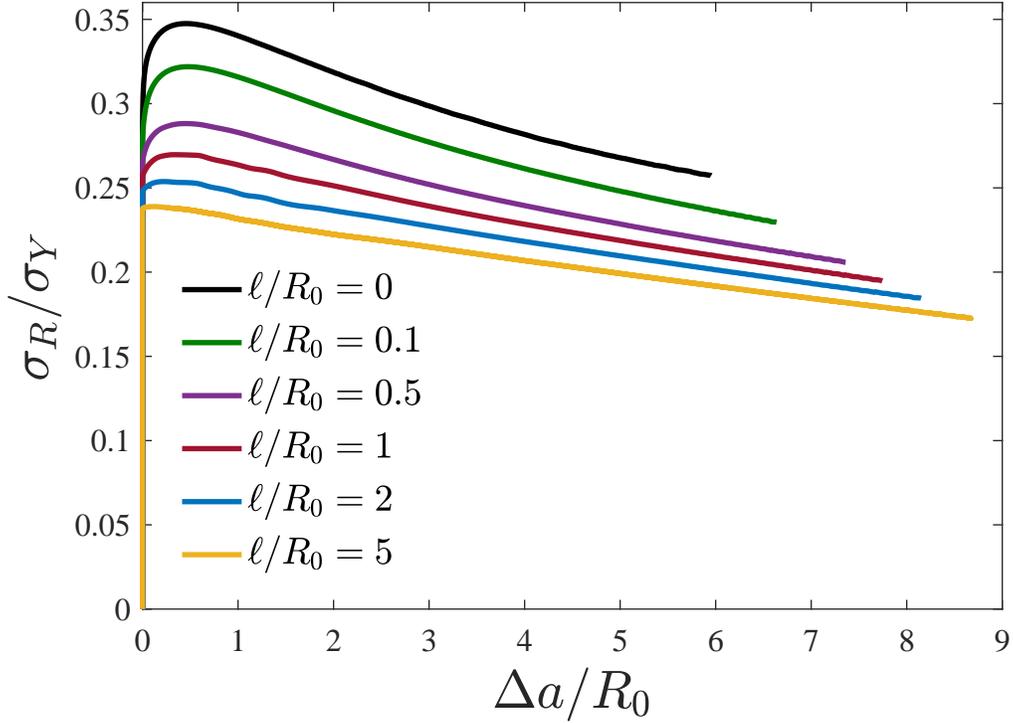}
\caption{Remote stress versus crack extension for the sharp V-notch case with $\alpha=60^{\circ}$. Results are shown for both conventional plasticity and MSG plasticity with different values of the length scale parameter. Material properties: $\sigma_Y/E=0.003$, $\nu=0.3$, $N=0.2$ and $\sigma_{max,0}=3.5 \sigma_Y$.}
\label{fig:StressCrack}
\end{figure}

The remote stress versus crack extension is shown in Fig. \ref{fig:StressCrack} for the same configuration. Here, $\sigma_R$ is obtained by measuring the vertical stress component in the element located in the upper left corner. In agreement with Fig. \ref{fig:LoadDisp}, the maximum remote stress that can be attained decreases significantly with augmenting $\ell$. Moreover, results reveal that the peak load at the outer boundary is reached at smaller crack sizes as the length parameter increases; this is due to the lower plastic dissipation that takes place. Hence, increasing the gradient contribution raises notch tip stresses, reducing the ductility and triggering fracture for lower values of the remote load.

\begin{figure}[H]
\centering
\includegraphics[scale=1.0]{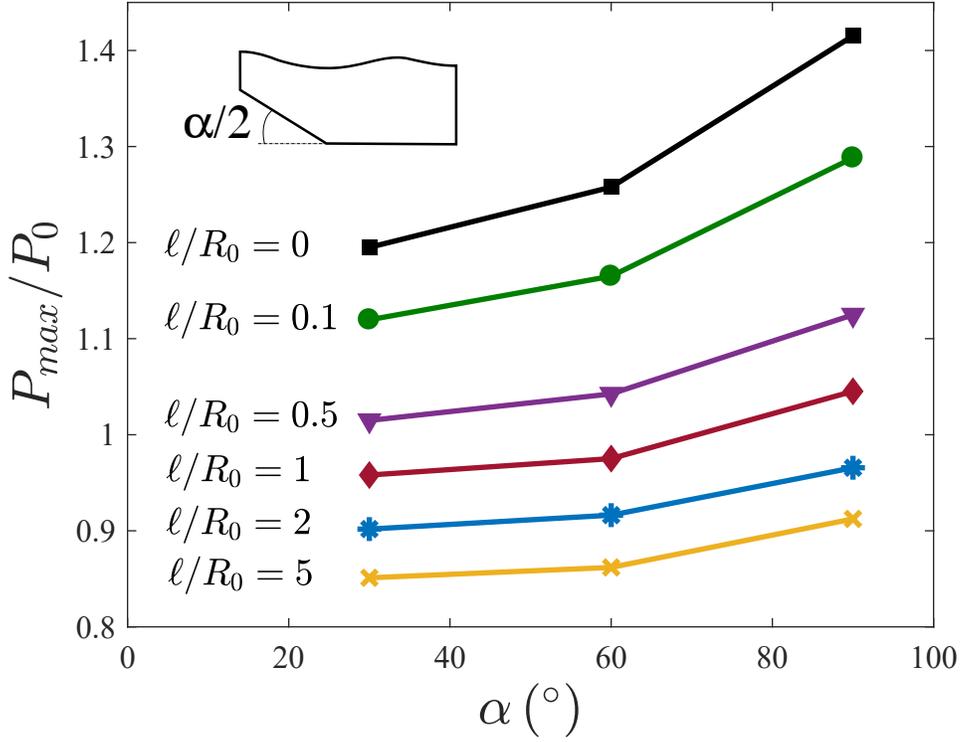}
\caption{Critical load versus notch angle for the sharp V-notch case. Results are shown for both conventional plasticity and MSG plasticity with different values of the length scale parameter. Material properties: $\sigma_Y/E=0.003$, $\nu=0.3$, $N=0.2$ and $\sigma_{max,0}=3.5 \sigma_Y$.}
\label{fig:MaxP}
\end{figure}

The influence of the plastic strain gradients on lowering the critical load in sharp V-notch specimens is quantified for three different angles. As shown in Fig. \ref{fig:MaxP}, as the notch angle decreases, the maximum force that can be attained decreases. This qualitative behavior can be easily understood from Fig. \ref{fig:CrackTipV} - higher notch tip stresses are attained with lower angles. A very strong gradient effect can be observed for the three cases; increasing $l/R_0$ increases the GND density, elevating the local stresses and lowering the critical force.

\begin{figure}[H]
\centering
\includegraphics[scale=1.0]{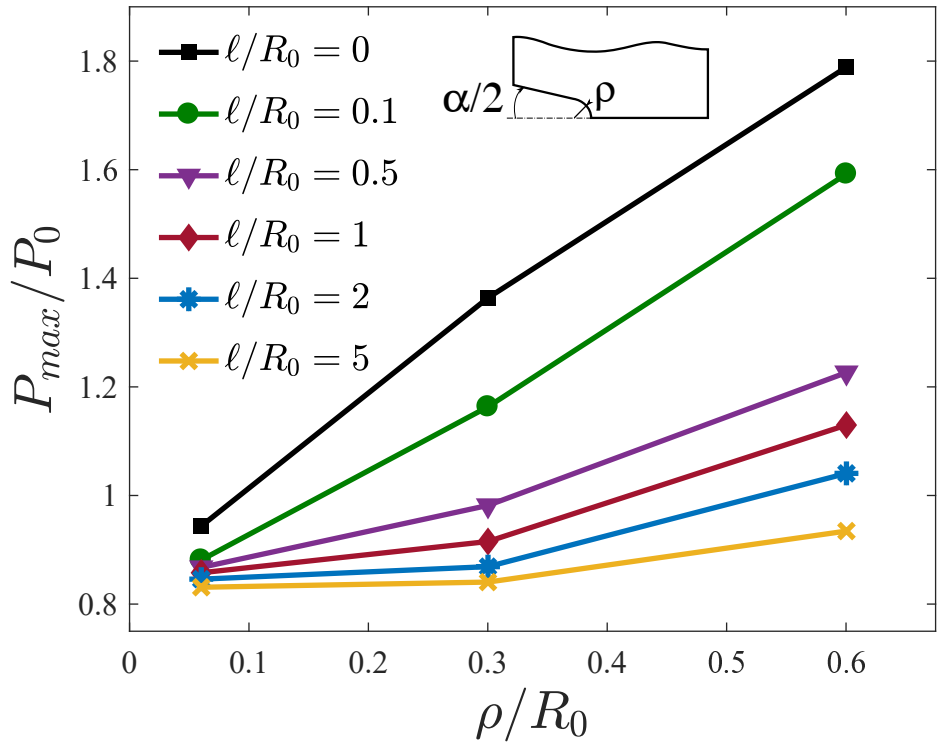}
\caption{Critical load versus notch radius for the blunted V-notch case for $\alpha=30^{\circ}$. Results are shown for both conventional plasticity and MSG plasticity with different values of the length scale parameter. Material properties: $\sigma_Y/E=0.003$, $\nu=0.3$, $N=0.2$ and $\sigma_{max,0}=2.5 \sigma_Y$.}
\label{fig:PvsRHOvb}
\end{figure}

The critical load is also computed for the blunted V-notch specimen for different notch radii and the same range of $\ell/R_0$ as in the sharp V-notch case; results are shown in Fig. \ref{fig:PvsRHOvb}. In agreement with the stationary notch tip stress calculations, lower $P_{max}$ values are attained by decreasing the notch radii. As in the sharp V-notch specimens, the GND effect persists for all the configurations examined. However, differences with conventional plasticity appear to be percentually higher for larger notch radii.  This is undoubtedly grounded on the fact that all calculations have been conducted for the same cohesive strength - for a given $\sigma_{max,0}$, failure takes place at lower load levels for smaller notch radii, and gradient effects decrease with the external load (not enough plasticity builds up, see \cite{IJSS2015,IJP2016}). Results are therefore sensitive to the choice of the cohesive strength.

\begin{figure}[H]
\centering
\includegraphics[scale=1.0]{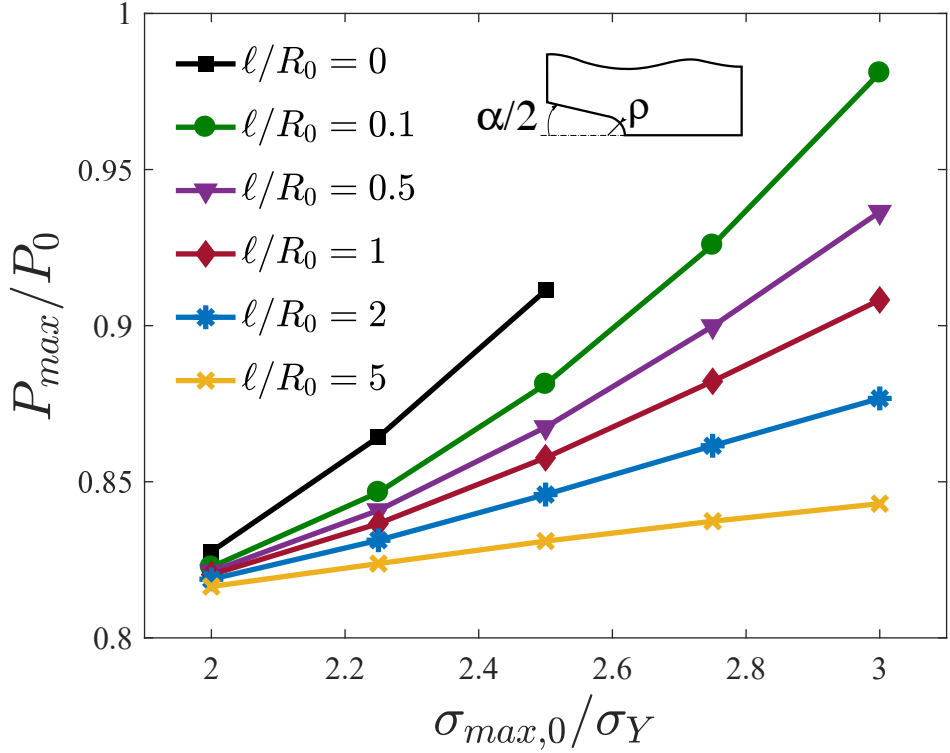}
\caption{Critical load versus cohesive strength for the blunted V-notch specimen with notch radius $\rho/\delta_n=15.8$ and $\alpha=30^{\circ}$. Results are shown for both conventional plasticity and MSG plasticity with different values of the length scale parameter. Material properties: $\sigma_Y/E=0.003$, $\nu=0.3$ and $N=0.2$.}
\label{fig:PvsSc}
\end{figure}

The variation of the maximum load with the cohesive strength for one particular notch radius is given in Fig. \ref{fig:PvsSc}. The length parameter and the reference load have been computed for each $\sigma_{max,0}/\sigma_Y$ to account for the influence of the cohesive strength on the fracture energy. As shown in the figure, higher critical loads are attained for larger values of $\sigma_{max,0}$ - the higher the cohesive strength, the more plastic dissipation contributes to the total energy release rate. Moreover, results show that differences between conventional and strain gradient plasticity increase significantly with increasing $\sigma_{max,0}$. Lower cohesive strengths can be attained for very small external loads, intrinsically associated with low levels of plastic deformation. Quantitative differences between conventional and gradient-enhanced constitutive relations are therefore sensitive to the value of $\sigma_{max,0}$. One should however note that the choice of cohesive strength is bounded by the maximum stress levels that can be attained with conventional plasticity. As shown in Fig. \ref{fig:PvsSc} no cracking is predicted for $\ell/R_0=0$ if $\sigma_{max}/\sigma_Y$ is higher than 2.5. From a physical viewpoint, it seems unlikely that an atomistically-grounded cohesive strength could be only 2.5 times the yield stress; accounting for the role of GNDs enables to consider more meaningful values. The quantitative differences reported between SGP and classic plasticity can therefore be substantially higher if $\sigma_{max}$ is increased.

\begin{figure}[H]
\centering
\includegraphics[scale=1.0]{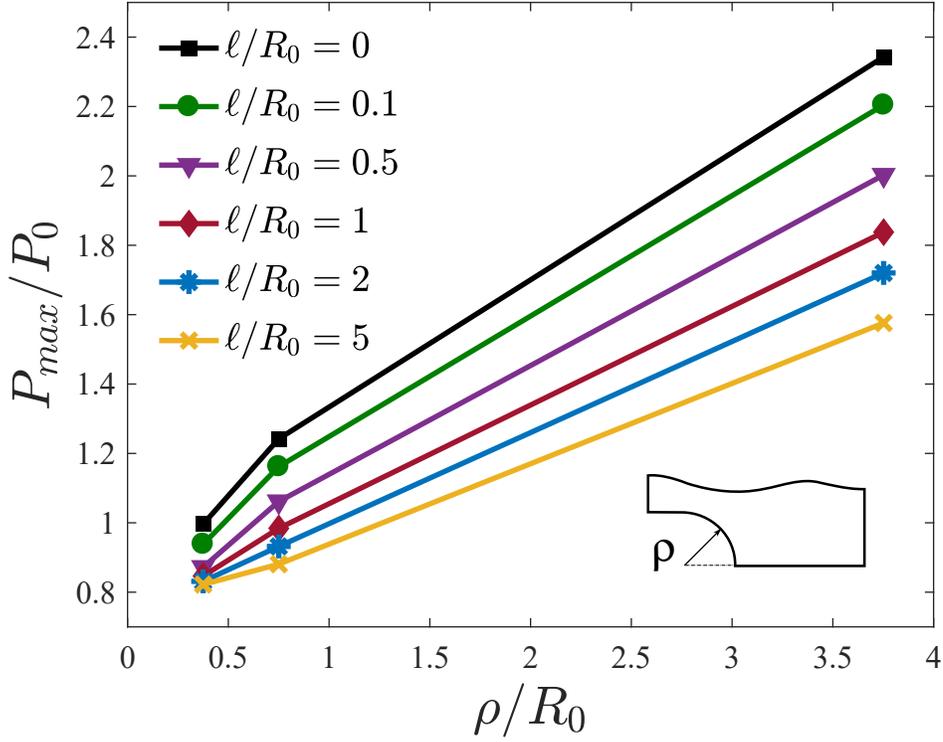}
\caption{Critical load versus notch radius for the U-notch case. Results are shown for both conventional plasticity and MSG plasticity with different values of the length scale parameter. Material properties: $\sigma_Y/E=0.003$, $\nu=0.3$, $N=0.2$ and $\sigma_{max,0}=2 \sigma_Y$.}
\label{fig:PvsRHOu}
\end{figure}

Finally, the peak load is computed for the U-notch case as a function of the notch radii and the length scale parameter. As shown in Fig. \ref{fig:PvsRHOu}, the maximum load increases with the notch radii, as reported in the blunted V-case. Important differences can be observed between classic and strain gradient plasticity formulations over the whole range of notch radii examined. Again, such differences seem to increase with the notch radii due to the larger loads involved.

\subsection{Cyclic loading}
\label{Sec:Cyclic}

We subsequently investigate notch-induced failure in the presence of cyclic loads. In order to do so we scale in time the external load by a sinusoidal function. The cyclic boundary conditions prescribed are characterized by the stress amplitude $\Delta \sigma = \sigma_{max} - \sigma_{min}$ and the stress ratio $R=\sigma_{min}/\sigma_{max}$. An initial prestressing is defined, such that the mean load equals the load amplitude, and both $R$ and $\Delta \sigma$ remain constant through the analysis. A stress ratio of $R=0.1$ is adopted throughout the study and, following \cite{Roe2003}, the accumulated cohesive length in (\ref{Eq:Damage}) is chosen to be $\delta_{\Sigma}=4 \delta_n$ and the endurance coefficient $\sigma_f / \sigma_{max,0}=0.25$. We use the same isotropic hardening law that has been used for the computation of the stationary notch tip fields and the cohesive crack propagation under monotonic loading. This choice comes at the cost of not being able to capture the Bauschinger effect displayed by many metallic materials under low load ratios. One should however note that our goal is to compare the responses obtained from classic and strain gradient plasticity theories under the same conventional hardening relation. Since gradient effects increase with plastic dissipation (see Sections \ref{Sec:Stationary} and \ref{Sec:Monotonic}), one would expect that the differences observed with isotropic hardening will be magnified if a kinematic hardening law is used. Taylor-based strain gradient plasticity models have been previously used with isotropic-like hardening laws to model fatigue in cracked components by Brinckmann and Siegmund \cite{Brinckmann2008,Brinckmann2008a}.\\

\begin{figure}[H]
\centering
\includegraphics[scale=1.05]{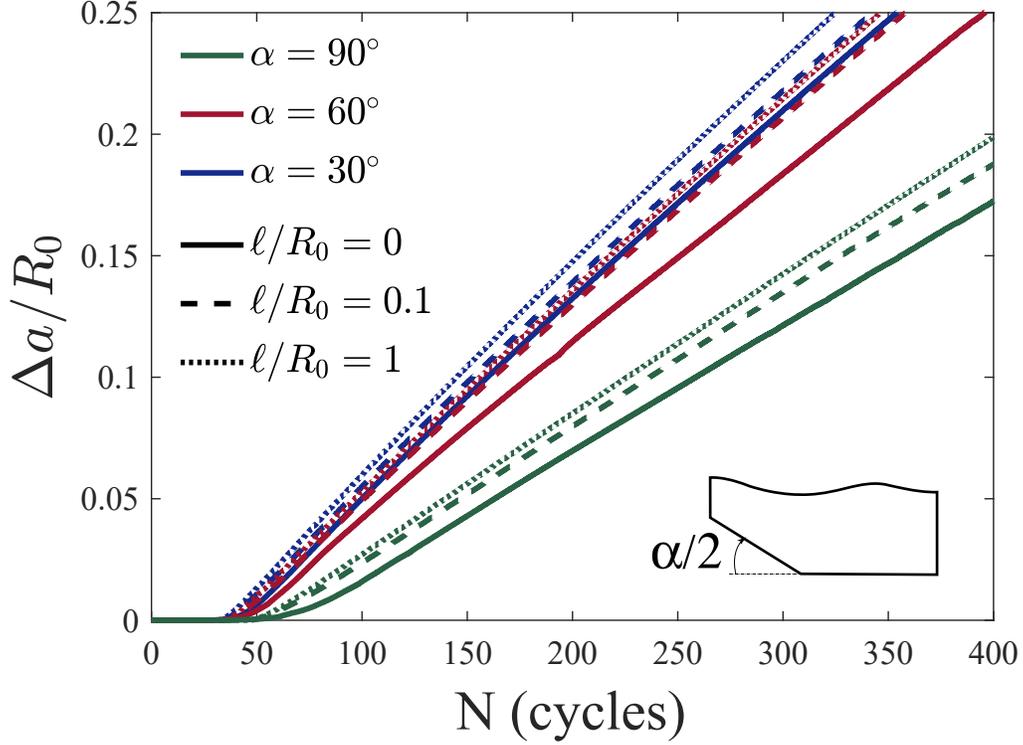}
\caption{Crack extension versus number of cycles for the sharp V-notch case with $\Delta \sigma/\sigma_Y=0.06$. Results are shown for different angles and different values of the length scale parameter. Problem parameters: $\sigma_Y/E=0.003$, $\nu=0.3$, $N=0.2$, $\sigma_{max,0}=3.75 \sigma_Y$ and $R=0.1$.}
\label{fig:AvsNv}
\end{figure}

Fig. \ref{fig:AvsNv} shows the crack extension in a sharp V-notched specimen as a function of the number of cycles for an stress amplitude of $\Delta \sigma / \sigma_Y=0.06$. Three notch angles have been considered, along with three different combinations of $\ell/R_0$. Relative to conventional plasticity predictions, SGP results show that: (i) cracking initiates before, and (ii) fatigue crack growth rates increase. These trends are observed in all the scenarios examined.

\begin{figure}[H]
\makebox[\linewidth][c]{%
        \begin{subfigure}[b]{0.6\textwidth}
                \centering
                \includegraphics[scale=0.61]{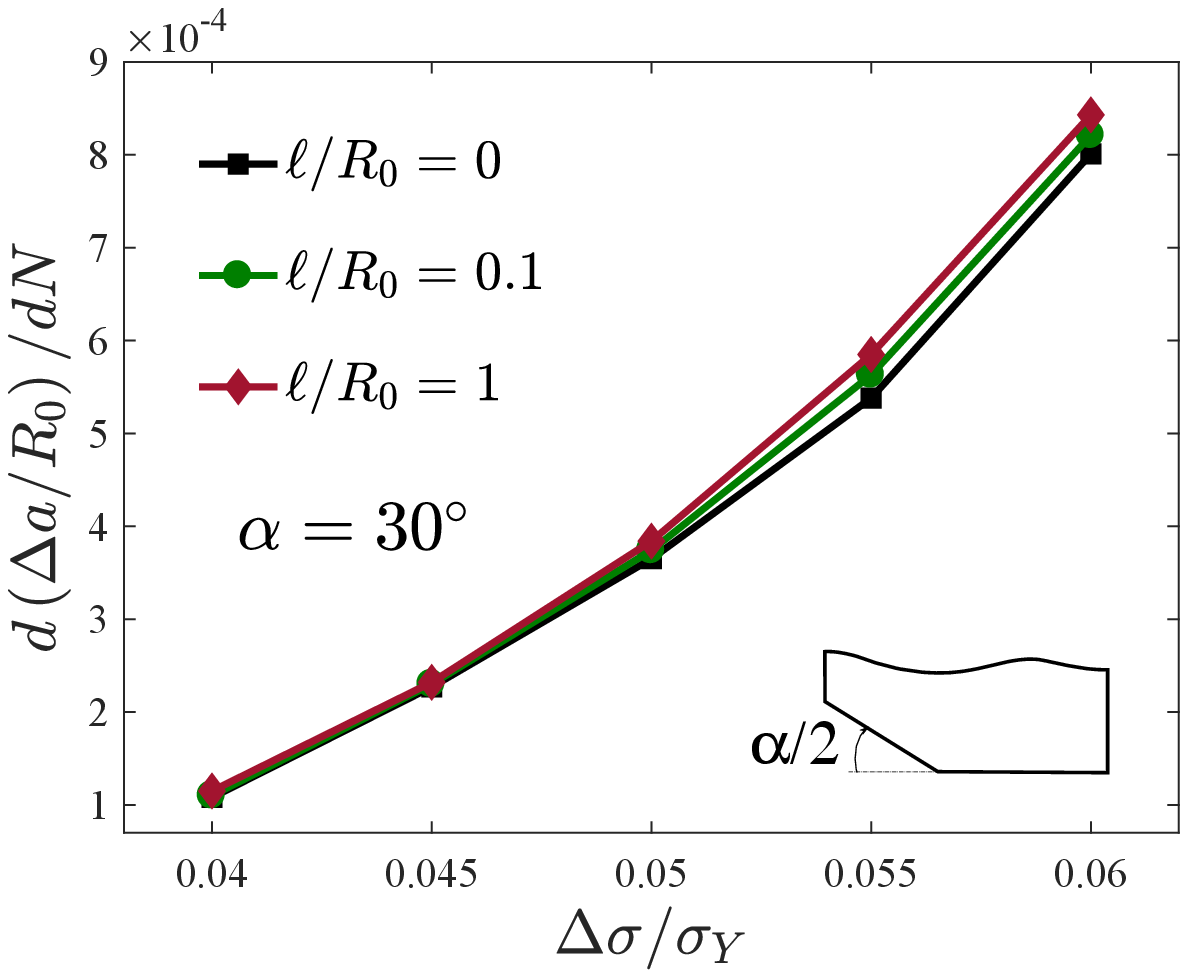}
                \caption{}
                \label{fig:FatigueV1}
        \end{subfigure}
        \begin{subfigure}[b]{0.6\textwidth}
                \raggedleft
                \includegraphics[scale=0.61]{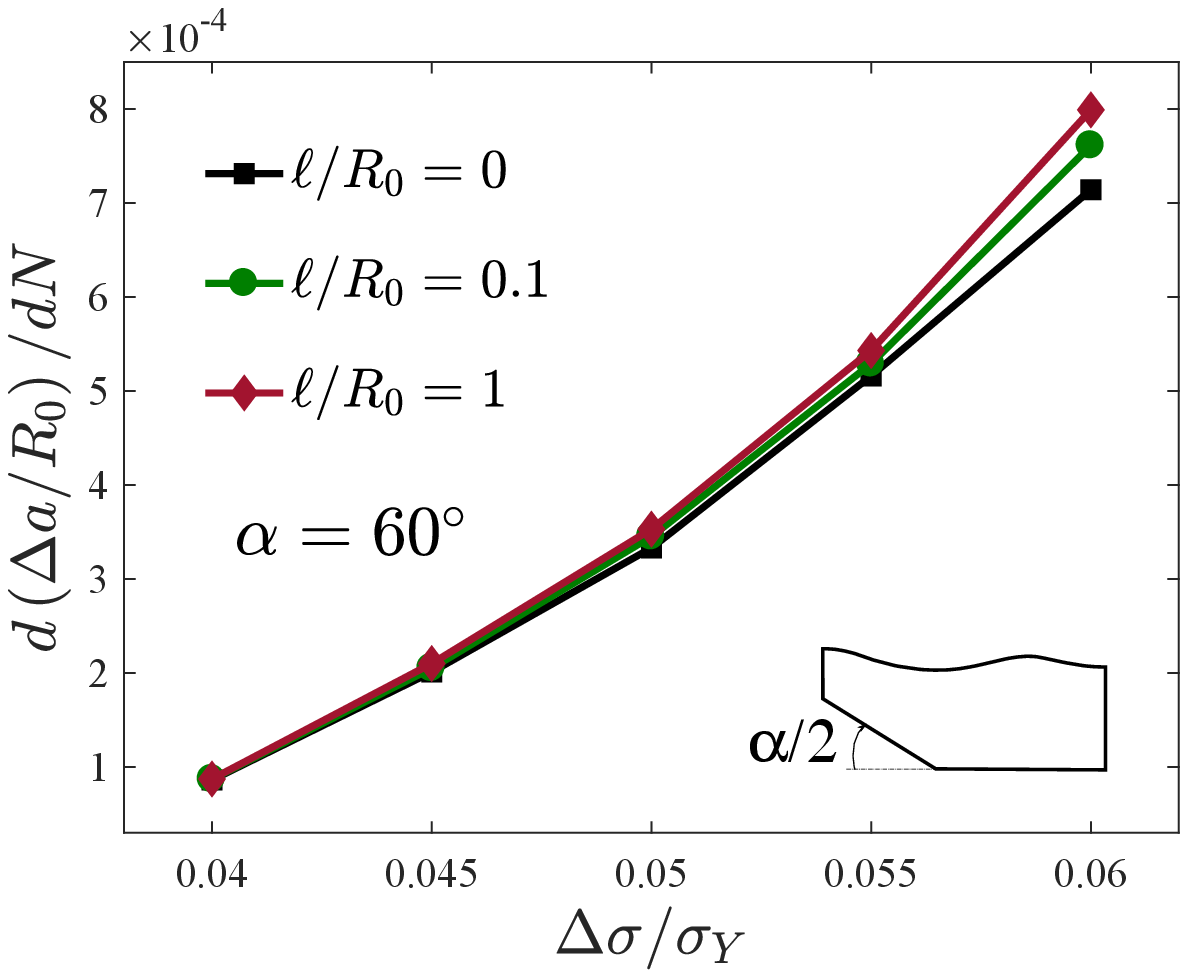}
                \caption{}
                \label{fig:FatigueV2}
        \end{subfigure}}

\makebox[\linewidth][c]{%
        \begin{subfigure}[b]{0.6\textwidth}
                \centering
                \includegraphics[scale=0.61]{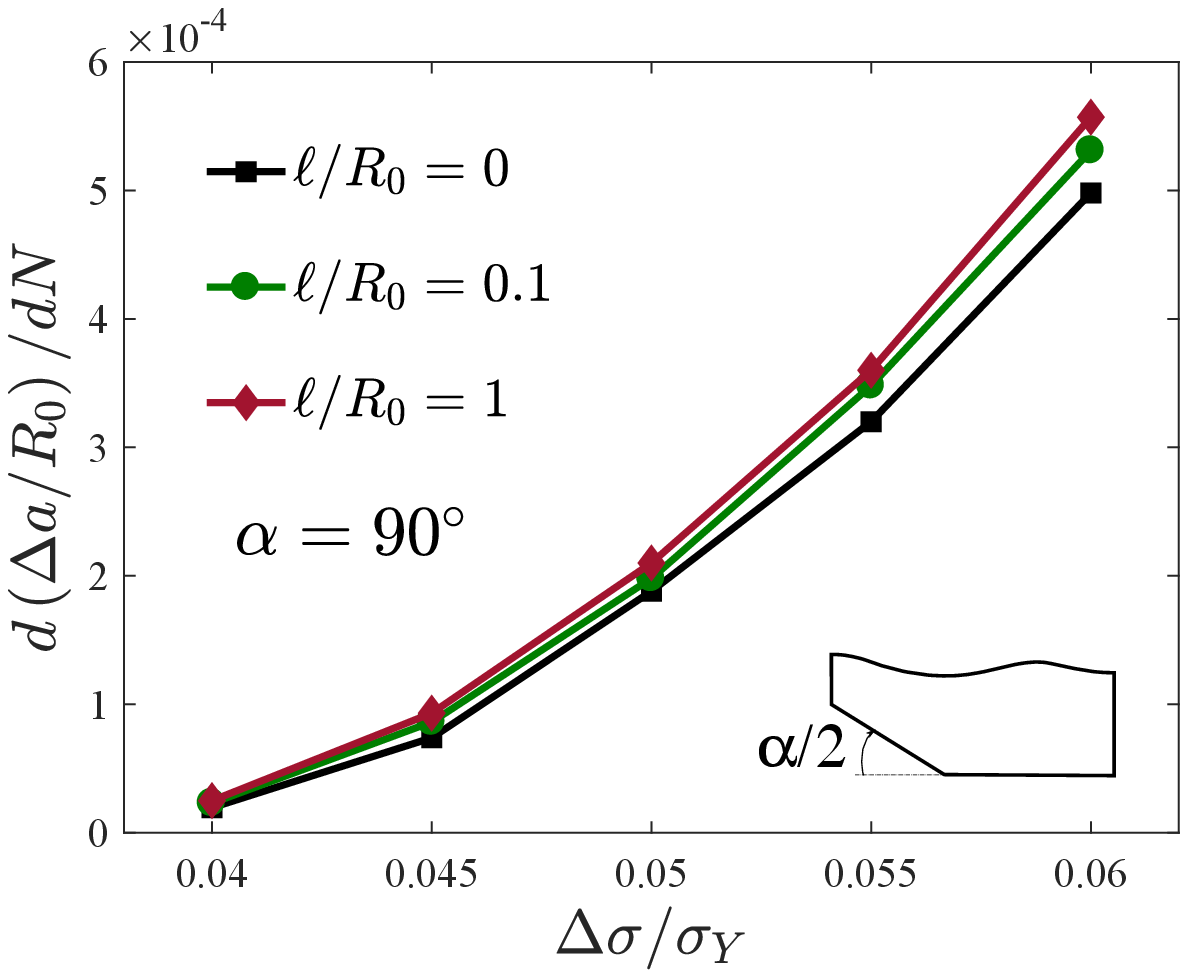}
                \caption{}
                \label{fig:FatigueV3}
        \end{subfigure}
        }       
        \caption{Fatigue crack growth rate versus stress amplitude for the sharp V-notch case with different notch angles: (a) $\alpha=30^{\circ}$, (b) $\alpha=60^{\circ}$, and (c) $\alpha=90^{\circ}$. Results are shown for different values of the length scale parameter. Problem parameters: $\sigma_Y/E=0.003$, $\nu=0.3$, $N=0.2$, $\sigma_{max,0}=3.75 \sigma_Y$ and $R=0.1$.}\label{fig:FatigueV}
\end{figure}

Fatigue crack growth rates are computed for a wide range of stress amplitudes and results are shown in Fig. \ref{fig:FatigueV}. In all cases an increase of the fatigue crack growth rates when increasing the external load and the length scale parameter can be observed. In addition, the GND-influence seems to increase with the external loads, although differences are not significant. Very little differences are in fact observed for the lower $\Delta \sigma/\sigma_Y$ bound, as cyclic damage reduces the cohesive strength and cracking takes place in the presence of considerable less plastic flow than in the monotonic case. As in Fig. \ref{fig:AvsNv}, fatigue crack growth rates increase as the notch angle decreases, for both conventional and gradient plasticity flow rules.

Cyclic crack propagation is also investigated for the blunted V-notched case. The results obtained in terms of crack extension as a function of the number of cycles are shown in Fig. \ref{fig:AvsNvb}. The finite element analysis reveals an increase on the fatigue crack growth rates and a decrease on the crack initiation cycle with augmenting $\ell/R_0$. It can also be observed that smaller notch radii lead to slightly higher fatigue crack growth rates.

\begin{figure}[H]
\centering
\includegraphics[scale=1.05]{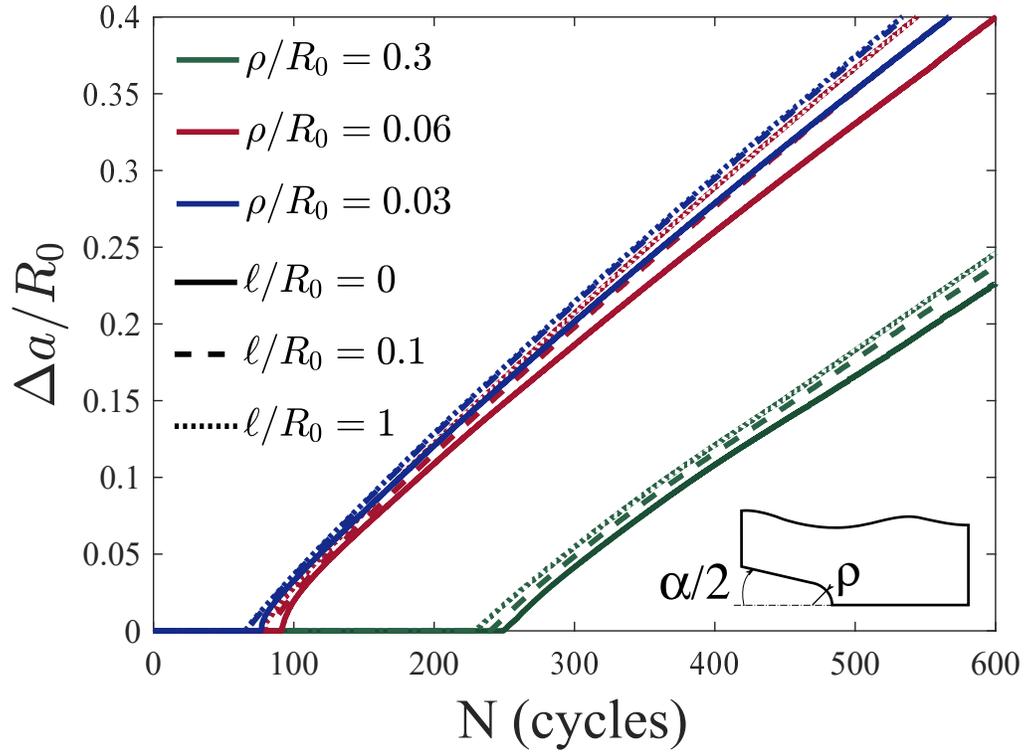}
\caption{Crack extension versus number of cycles for the blunted V-notch case with $\Delta \sigma/\sigma_Y=0.04$. Results are shown for different notch radii and different values of the length scale parameter. Problem parameters: $\sigma_Y/E=0.003$, $\nu=0.3$, $N=0.2$, $\sigma_{max,0}=2.5 \sigma_Y$ and $R=0.1$.}
\label{fig:AvsNvb}
\end{figure}

\begin{figure}[H]
\makebox[\linewidth][c]{%
        \begin{subfigure}[b]{0.6\textwidth}
                \centering
                \includegraphics[scale=0.61]{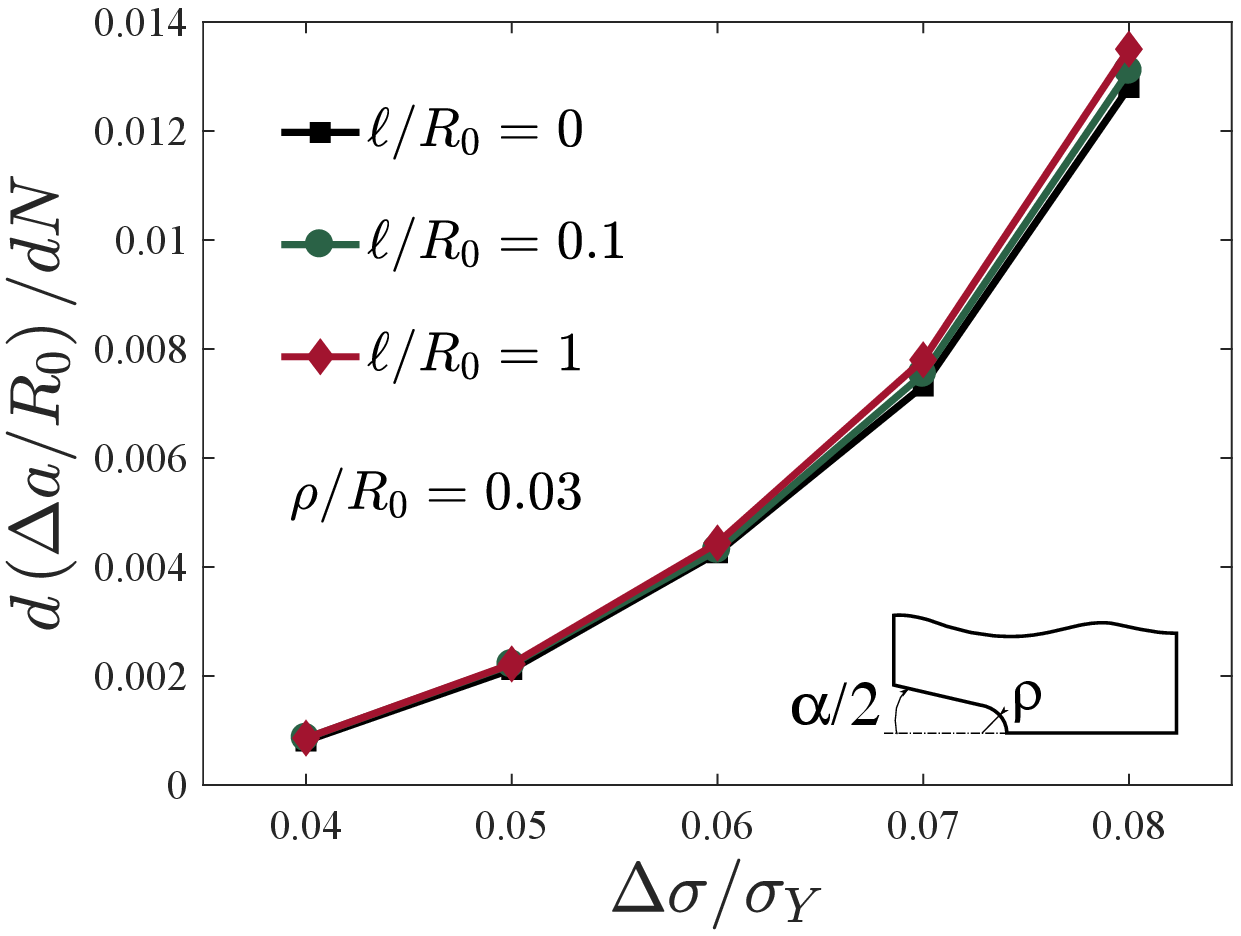}
                \caption{}
                \label{fig:FatigueVb1}
        \end{subfigure}
        \begin{subfigure}[b]{0.6\textwidth}
                \raggedleft
                \includegraphics[scale=0.61]{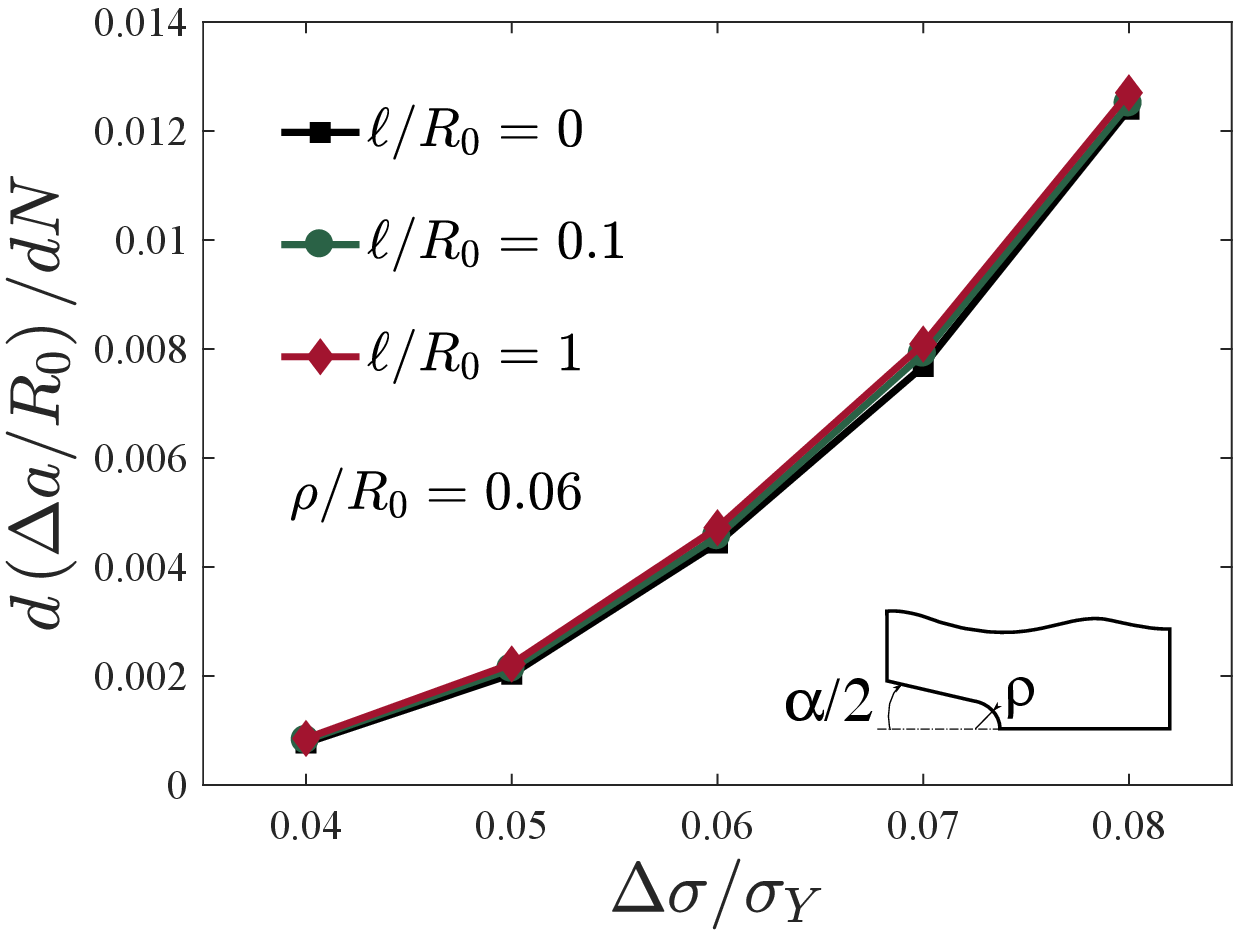}
                \caption{}
                \label{fig:FatigueVb2}
        \end{subfigure}}

\makebox[\linewidth][c]{%
        \begin{subfigure}[b]{0.6\textwidth}
                \centering
                \includegraphics[scale=0.61]{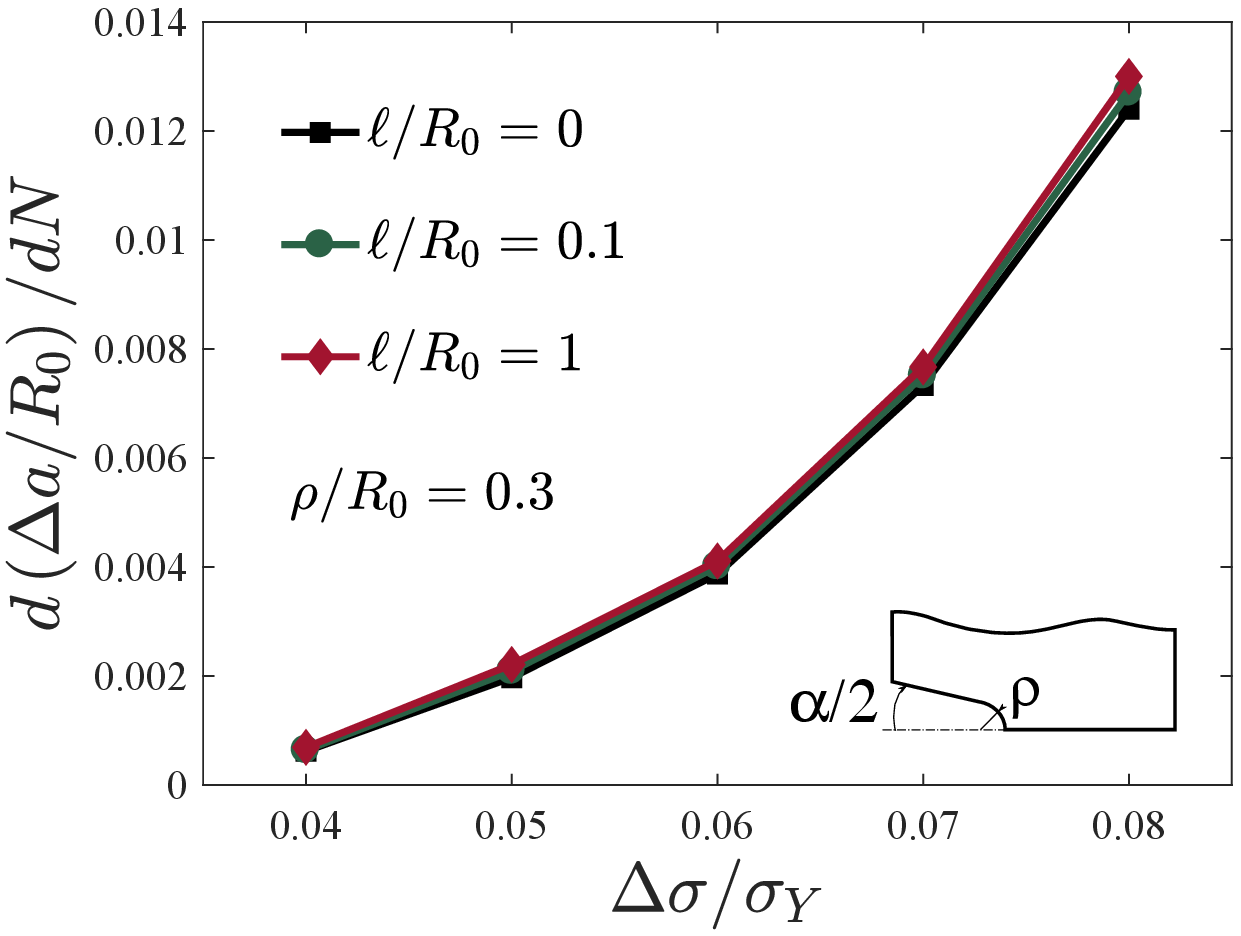}
                \caption{}
                \label{fig:FatigueVb3}
        \end{subfigure}
        }       
        \caption{Fatigue crack growth rate versus stress amplitude for the blunted V-notch case with different notch radii: (a) $\rho/R_0=0.03$, (b) $\rho/R_0=0.06$, and (c) $\rho/R_0=0.3$. Results are shown for different values of the length scale parameter. Problem parameters: $\sigma_Y/E=0.003$, $\nu=0.3$, $N=0.2$, $\sigma_{max,0}=2.5 \sigma_Y$ and $R=0.1$.}\label{fig:FatigueVb}
\end{figure}

Fig. \ref{fig:FatigueVb} quantifies fatigue crack growth rates as a function of the external load for the blunted V-notch specimens. Results show very little sensitivity to the notch radii. This also holds true for the GND-effect, which seems to be much more sensitive to the external load rather than the geometry; differences with conventional plasticity increase as $\Delta \sigma$ increases.\\

Similar qualitative trends are observed for the U-notch geometry. Fig. \ref{fig:AvsNvb} shows the crack extension versus the number of loading cycles for three notch radii and three length scale parameters. Again, the number of cycles required to initiate cracking reduces with larger $\ell/R_0$ and smaller notch radii, and gradient effects translate into an increase of the fatigue crack growth rates.

\begin{figure}[H]
\centering
\includegraphics[scale=1.05]{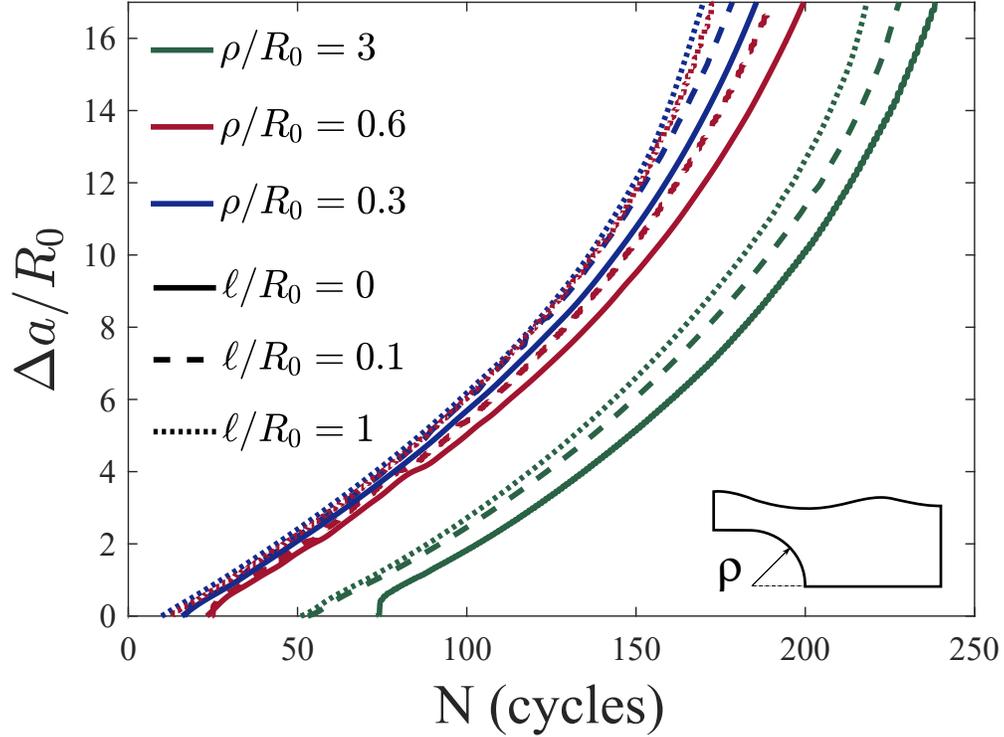}
\caption{Crack extension versus number of cycles for the U-notch case with $\Delta \sigma/\sigma_Y=0.13$. Results are shown for different notch radii and different values of the length scale parameter. Problem parameters: $\sigma_Y/E=0.003$, $\nu=0.3$, $N=0.2$, $\sigma_{max,0}=2.5 \sigma_Y$ and $R=0.1$.}
\label{fig:AvsNvb}
\end{figure}

\begin{figure}[H]
\makebox[\linewidth][c]{%
        \begin{subfigure}[b]{0.6\textwidth}
                \centering
                \includegraphics[scale=0.61]{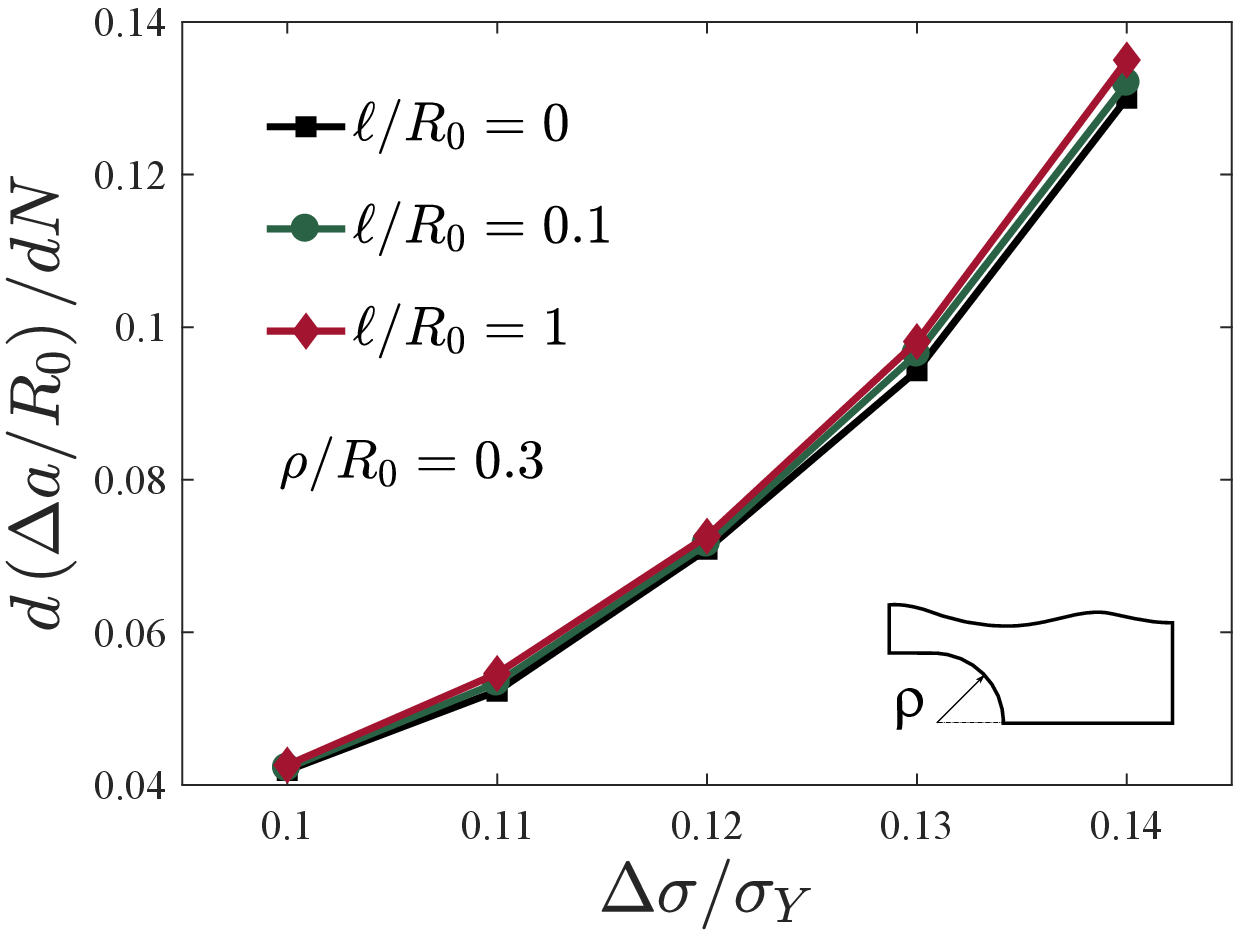}
                \caption{}
                \label{fig:FatigueVb1}
        \end{subfigure}
        \begin{subfigure}[b]{0.6\textwidth}
                \raggedleft
                \includegraphics[scale=0.61]{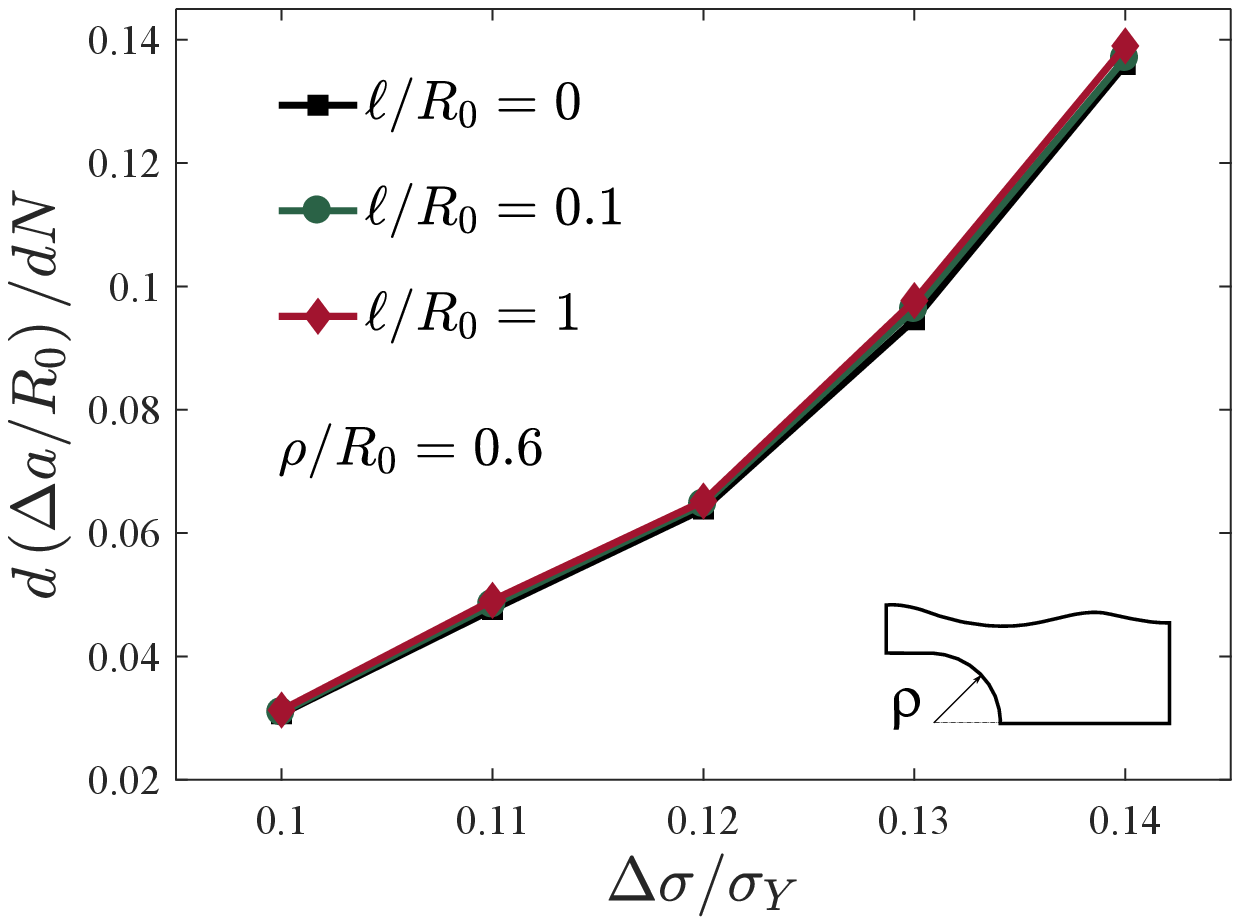}
                \caption{}
                \label{fig:FatigueVb2}
        \end{subfigure}}

\makebox[\linewidth][c]{%
        \begin{subfigure}[b]{0.6\textwidth}
                \centering
                \includegraphics[scale=0.61]{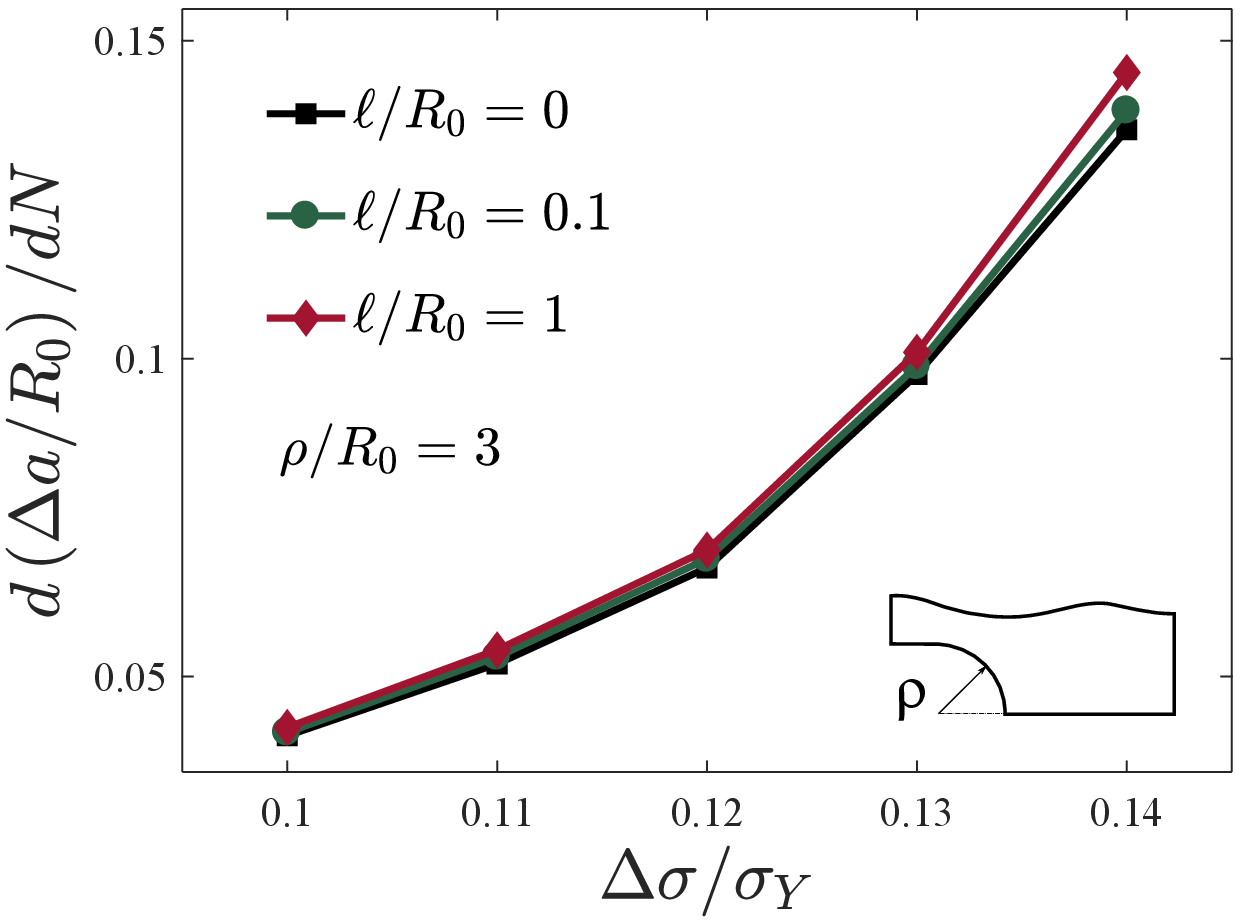}
                \caption{}
                \label{fig:FatigueVb3}
        \end{subfigure}
        }       
        \caption{Fatigue crack growth rate versus stress amplitude for the U-notch case with different notch radii: (a) $\rho/R_0=0.3$, (b) $\rho/R_0=0.6$, and (c) $\rho/R_0=3$. Results are shown for different values of the length scale parameter. Problem parameters: $\sigma_Y/E=0.003$, $\nu=0.3$, $N=0.2$, $\sigma_{max,0}=2.5 \sigma_Y$ and $R=0.1$.}\label{fig:FatigueU}
\end{figure}

Normalized fatigue crack growth rates $da/dN$ in U-notched specimens are shown as a function of the stress ratio in Fig. \ref{fig:FatigueU}. Differences between SGP and conventional plasticity increase with the external load, as in the sharp and blunted V-notched cases. Results show nevertheless little sensitivity, particularly for smaller stress amplitudes.

\section{Conclusions}
\label{Concluding remarks}

The first investigation on the role of plastic strain gradients in notched assisted failure has been presented. The influence of geometrically necessary dislocations (GNDs) in elevating the stresses ahead of the notch tip is thoroughly examined by means of a mechanism-based strain gradient plasticity theory. A total of 9 different geometries have been considered from the most common notch types: sharp V (with 3 angles), blunted V (with 3 radii) and U (with 3 radii). A comprehensive finite element investigation has been conducted including the analysis of stationary notch tip stresses, and crack propagation under monotonic and cyclic loading. A suitable cohesive zone formulation has been employed for the latter, which includes a cycle-dependent traction-separation relation. Results reveal that GNDs have a strong impact on the failure of notched components. Particularly, the following aspects must be highlighted:

\begin{itemize}
\item Large strain gradients in the vicinity of the notch promote local hardening and lead to notch tip stresses that much larger than those predicted by means of conventional plasticity.
\item Smaller notches show a very significant gradient-enhanced stress elevation over a micron-scale physical length; as opposed to larger notches, which lead to a larger gradient-dominated region with a lesser stress rise.
\item Monotonic crack propagation studies show that GNDs bring a substantial reduction on the ductility and the maximum carrying capacity.
\item Under cyclic loading, gradient effects translate into a noticeable enhancement of fatigue crack growth rates and a premature initiation of cracking.
\end{itemize}

Non-local strain gradient modeling of notch-induced structural integrity appears therefore indispensable to obtain high fidelity predictions in metallic components.
 
\section{Acknowledgments}
\label{Acknowledge of funding}

The authors gratefully acknowledge financial support from the Ministry of Economy and Competitiveness of Spain through grant MAT2014-58738-C3. E. Mart\'{\i}nez-Pa\~neda also acknowledges financial support from the People Programme (Marie Curie Actions) of the European Union's Seventh Framework Programme (FP7/2007-2013) under REA grant agreement n$^{\circ}$ 609405 (COFUNDPostdocDTU). Javier Segurado (UPM, IMDEA Materials) is acknowledged for helpful discussions relative to the control algorithm.





\bibliographystyle{elsarticle-num} 
\bibliography{library}

\end{document}